\documentclass[a4paper, 11 pt]{article}  
\usepackage{fullpage}
\usepackage{setspace}
\usepackage{color}
\usepackage[]{graphicx}
\usepackage{epsfig} 
\usepackage{amsmath}
\usepackage{amssymb}
\usepackage[mathscr]{eucal}
\usepackage{bm}
\usepackage{subfigure}
\usepackage{theorem}

\pdfoutput=1

\newcommand{\beq}{\begin{equation}}
\newcommand{\eeq}{\end{equation}}
\newcommand{\beqa}{\begin{eqnarray}}
\newcommand{\eeqa}{\end{eqnarray}}
\newcommand{\beqan}{\begin{eqnarray*}}
\newcommand{\eeqan}{\end{eqnarray*}}

\usepackage{xr}

{\theorembodyfont{\upshape}

}

\title{The geometric phase of stock trading}

\author{C. Altafini \\
Division of Automatic Control, Dept. of Electrical Engineering, \\
Link\"oping University, SE-58183, Link\"oping, Sweden. \\ email: {\tt\small claudio.altafini@liu.se}}

\begin{document}

\maketitle

\begin{abstract}
Geometric phases describe how in a continuous-time dynamical system the displacement of a variable (called phase variable) can be related to other variables (shape variables) undergoing a cyclic motion, according to an area rule. 
The aim of this paper is to show that geometric phases can exist also for discrete-time systems, and even when the cycles in shape space have zero area. 
A context in which this principle can be applied is stock trading. 
A zero-area cycle in shape space represents the type of trading operations normally carried out by high-frequency traders (entering and exiting a position on a fast time-scale), while the phase variable represents the cash balance of a trader. 
Under the assumption that trading impacts stock prices, even zero-area cyclic trading operations can induce geometric phases, i.e., profits or losses, without affecting the stock quote. 
\end{abstract}

%
%

\section{Introduction}

Geometric phases are ``cycles that effect changes'' \cite{Brockett1997Cycle}. They appear in systems in which a cyclic change in some of the variables (called {\em shape} variables) induces a non-zero net motion on other variables (called {\em phase} variables). 
They are well studied phenomena for instance in classical and quantum mechanics \cite{Berry199034,Marsden2,wilczek1989geometric}, molecular systems, \cite{Meald1992Geometric}, robotics \cite{Laumond6,murray1994mathematical} and control theory \cite{Bloch4}. 
They explain how a falling cat can manage to land always on its feet \cite{Montgomery1993Gauge}, how a bacterium can propel itself in a highly viscous fluid \cite{Shapere1989Geometry}, and how we can parallel park a car \cite{MurrayS1}, see \cite{Batterman2003527} for an overview.

\begin{figure}[t!]
\begin{center}
\subfigure[]{
\includegraphics[angle=0, trim=2cm 2cm 3cm 3cm, clip=true,width=7cm]{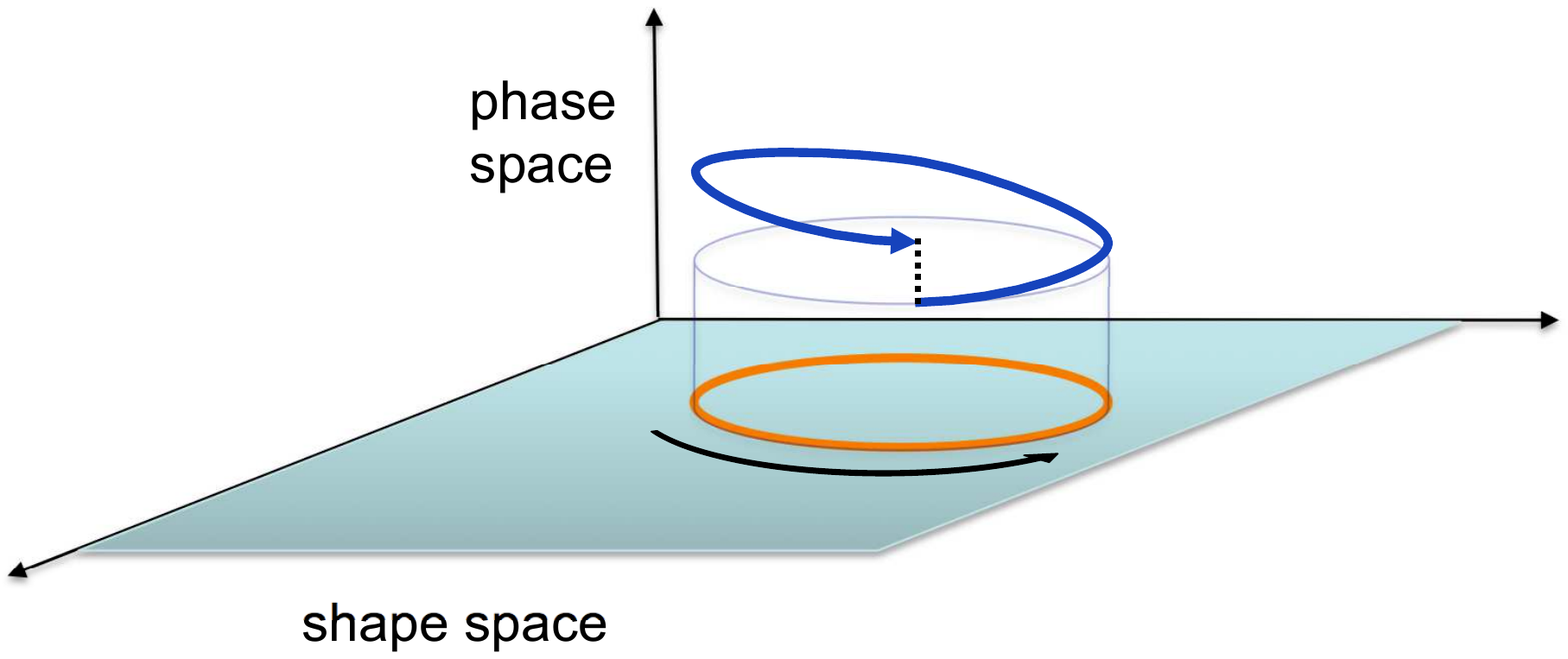}} $ \qquad $ 
\subfigure[]{
\includegraphics[angle=0, trim=2cm 2cm 3cm 3cm, clip=true,width=7cm]{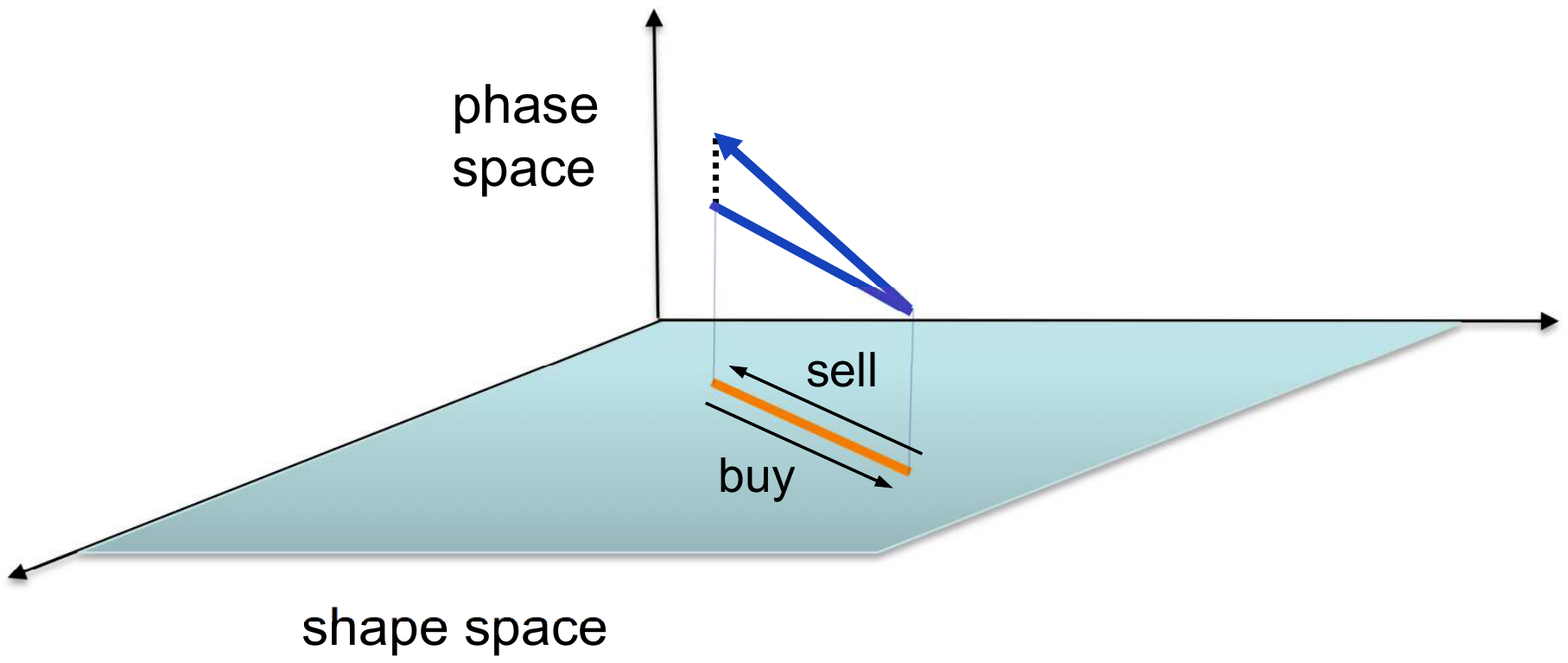}} 
\caption[]{Geometric phase in continuous and discrete time. Panel (a): classical picture for geometric phase of a continuous-time dynamical system: a cyclic path in shape space (orange curve) induces a net motion on the phase variable (blue curve), proportional to the area of the cycle in shape space, see Appendix for an explicit model. Panel (b): geometric phase for a discrete-time dynamical system. A cyclic trajectory, even of zero-area in shape space (orange), induces a net nonzero motion in the phase variable (blue).
In the stock trading application, the shape variables are the amount of stocks owned by a trader and the stock quote. 
The phase variable is the cash balance of the trader.}
\label{fig:geometric_phase}
\end{center}
\end{figure}

Originally introduced in quantum physics by Berry \cite{Berry45}, they are normally used in conjunction with nonintegrable systems of continuous-time differential equations.
When a cyclic trajectory is generated in shape space, nonintegrability of the ODEs induces a motion which is non-periodic in phase space, see Fig.~\ref{fig:geometric_phase} (a). 
The amplitude of this phase displacement is proportional to the area of the cyclic path in shape space. 
In particular, a zero-area cycle yields no geometric phase. 

There is however a case in which even a zero-area cycle can induce a nontrivial geometric phase, and it is when we consider difference equations in discrete time. 
Some of the peculiarities of nonintegrable discrete-time dynamical systems have been known since several decades in control theory \cite{Jakubczyk1990Controllability,Monaco1986}.
However, in our knowledge, the properties of discrete time geometric phases have never been investigated, let alone the existence of phase motions induced by zero-area shape cycles.

The aim of this paper is to shown that indeed such geometric phases exist, and to present an application where they admit a useful interpretation.
The application we want to discuss is stock trading. 
In the idealized stock exchange we consider, we can take as shape variables the amount of a certain stock owned by a trader and the quote of that stock. 
The discrete events which constitute the time instants of interest for our model are the trading events that concern the stock we are dealing with. 
Using a discrete-time dynamics for them is natural, especially when fast time scales are considered. 
The phase variable is the cash balance of our trader, which is a function of how many stocks are bough/sold but also of the quote at which they are bought/sold. 
The product of these two quantities gives the nonlinearity required to have a nonintegrable system of difference equations. 
Assume that our trader is the only actor in our stock market, and that his market orders are executed against a limit order book \cite{Gould2013Limit,cartea2015algorithmic}. 
Assume further that trading impacts prices, i.e., a buy order drives prices up, while a sell order drives prices down. 
When nothing else happens to our stock, a zero-area cyclic trajectory in shape space is a sequence of a buy order followed by a sell order of equal magnitude, see Fig.~\ref{fig:geometric_phase} (b).
At the end of the cycle, the stock quote (a shape variable) is unchanged, as the two price impacts compensate each other. 
However the cash balance of the trader (the phase variable) is {\em not} zero at the end of the cycle.
In fact, the quote at which the stocks are bought does not contain the price variation due to the action of buying itself.
However, this price impact is incorporated in the quote at selling time, which makes the quote at which stocks are sold higher than that at which the stocks are bought, as an effect of the act of trading itself. 
This is the geometric phase of stock trading. 

In what follows an idealized scenario is presented first, as a way to shown that nontrivial geometric phases for zero-area shape trajectories indeed exist.  
The scenario is then rendered more realistic by incorporating in the model a bid/ask spread for the stock quote, commission fees, and a stock quote drift (representing the operations of all other traders of the same stock), without altering the nature of the geometric phase phenomenon just described. 

Entering and liquidating a position, possibly very quickly, is the typical pattern of high-frequency trading (i.e., fast computerized trading with short holding periods and no inventory \cite{duhigg_2009,aldridge2013high,Brogaard14052014,Carrion2013680}). 
It is shown in the paper that practices like front-running, sometimes attributed to high-frequency trading \cite{lewis2014flash}, consist in interlacing fast cycles with operations of slower ``classical'' traders, and are essentially a way of increasing the price impact (and hence the geometric phase) during a cycle in shape space.

\section{Models}

Consider an idealized ``stock exchange'' scenario consisting of a single trader trading on a single stock through market orders, executed against a limit order book (\cite{Gould2013Limit}, more details are given below). 
Denote $ y(t) $ the quantity of stocks owned by the trader at time $ t$ and with $ s(t)$ the stock quote.
When the trader buys or sell stocks, his cash balance $ z(t)$ changes accordingly.

We make the assumption that trades impact prices: selling stocks drives prices down, while buying stocks drives prices up.
In discrete time an elementary model for the scenario being considered is the following: 

\beqa
y(t_{i+1}) & = & y(t_i) + u(t_i) \label{eq:basicmodel1} \\
s(t_{i+1}) & = & s(t_i) + r u(t_i) \label{eq:basicmodel2} \\
z(t_{i+1}) & = & z(t_i) - s(t_i) u(t_i) \label{eq:basicmodel3} .
\eeqa
Here $ u(t_i) = y(t_{i+1}) - y(t_i) $ is the amount of stock being bought ($ u(t_i)>0 $) or sold ($u(t_i) <0$) at time $ t_i$ by the trader. 
At time of execution, a buy order $ (u(t_i) >0$) raises the quote of the stock, i.e., $ s(t_{i+1}) > s(t_i)$, while the cash balance decreases $ z(t_{i+1}) < z(t_i) $. 
The events represented by \eqref{eq:basicmodel1}--\eqref{eq:basicmodel3} all have the same time stamp $ t_i$, meaning that for all 3 equations the variables are updated at the next time instant, $ t_{i+1}$. The length of the time interval $ \Delta t_i = t_{i+1} - t_i $ can be very small, provided that $ \Delta t_i>0$. 
The coefficient $ r$ quantifies the price impact, i.e., how influential an order is for a certain stock. 
Its value (here assumed constant) may depend on the total volume of the stock, its liquidity, volatility, etc \cite{Cont07062013}, see also below for a more general formula.
In vector form, calling $ x(t) = \begin{bmatrix} y(t) \\ s(t)  \\ z(t) \end{bmatrix} $ the state of the system, the equations \eqref{eq:basicmodel1}--\eqref{eq:basicmodel3} can be written as
\beq
x(t_{i+1}) 
= x(t_i ) + \begin{bmatrix} 1  \\ r  \\ - s(t_i)  \end{bmatrix} u(t_i ).
\label{eq:basicmodel_vect}
\eeq
In \eqref{eq:basicmodel_vect}, $ u(t) $ can be considered as the input of the system.

\begin{figure}[h!]
\begin{center}
\subfigure[]{
\includegraphics[width=5.5cm]{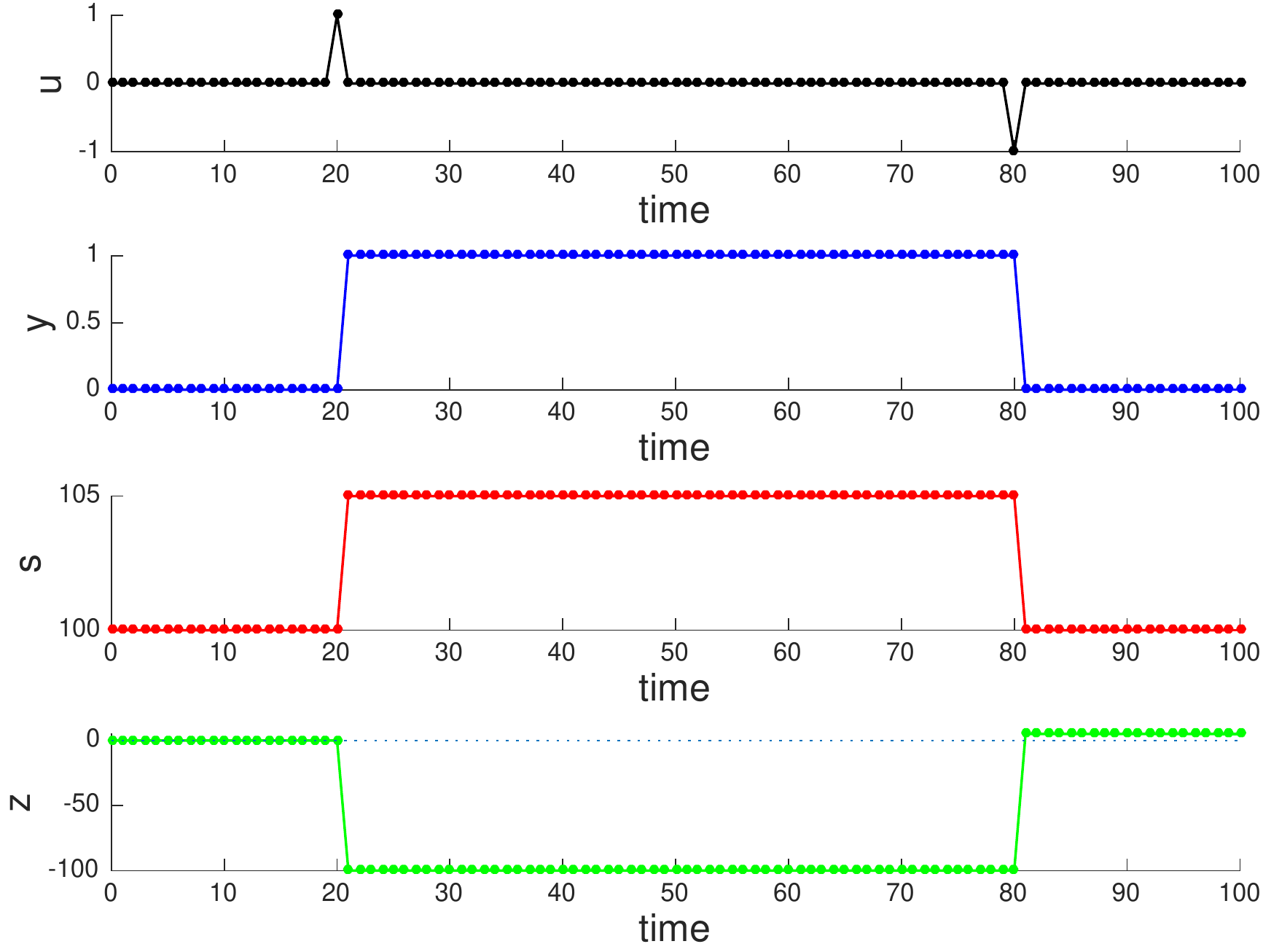}} 
\subfigure[]{
\includegraphics[width=5.5cm]{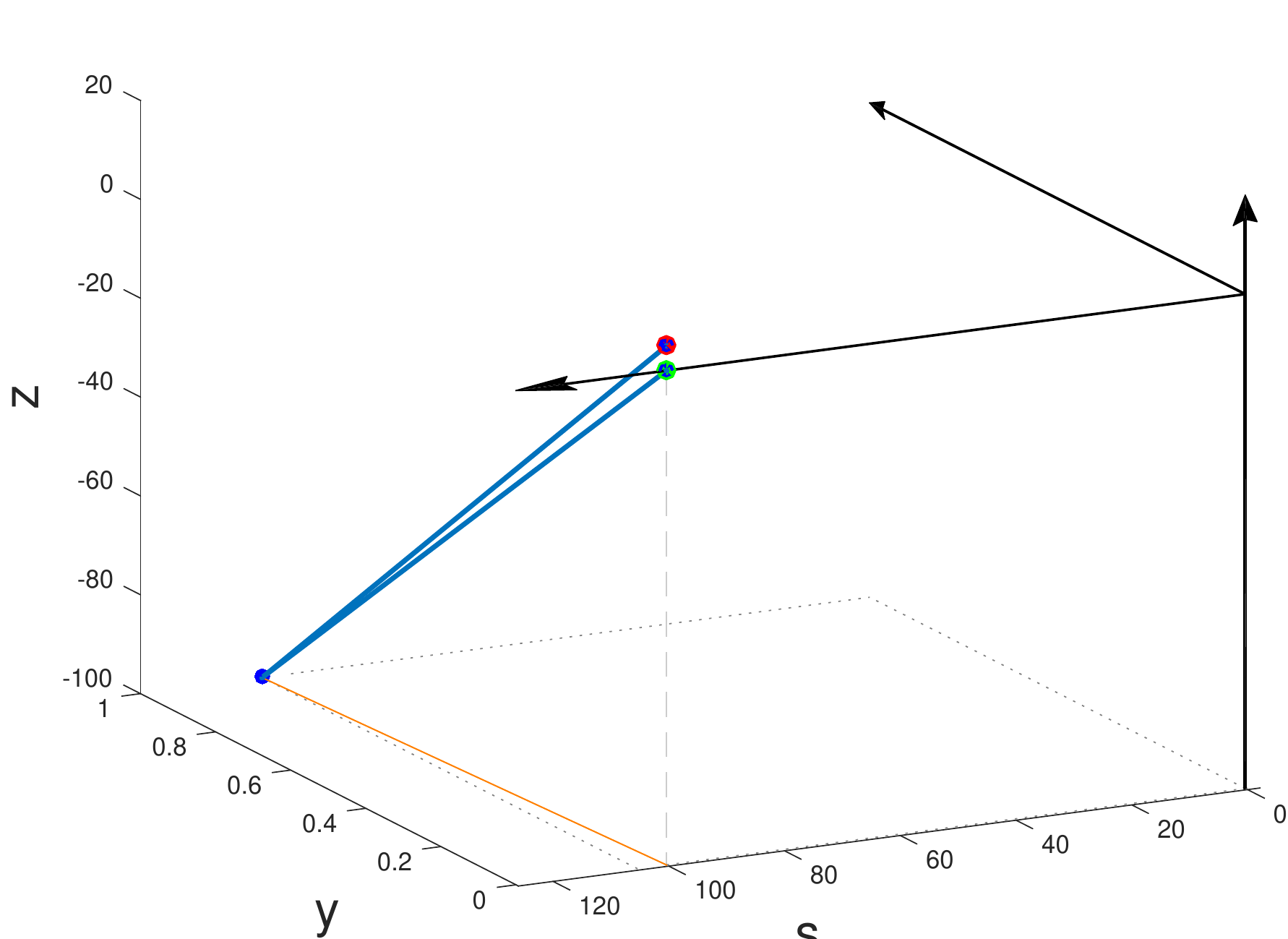}}
\subfigure[]{
\includegraphics[width=5.5cm]{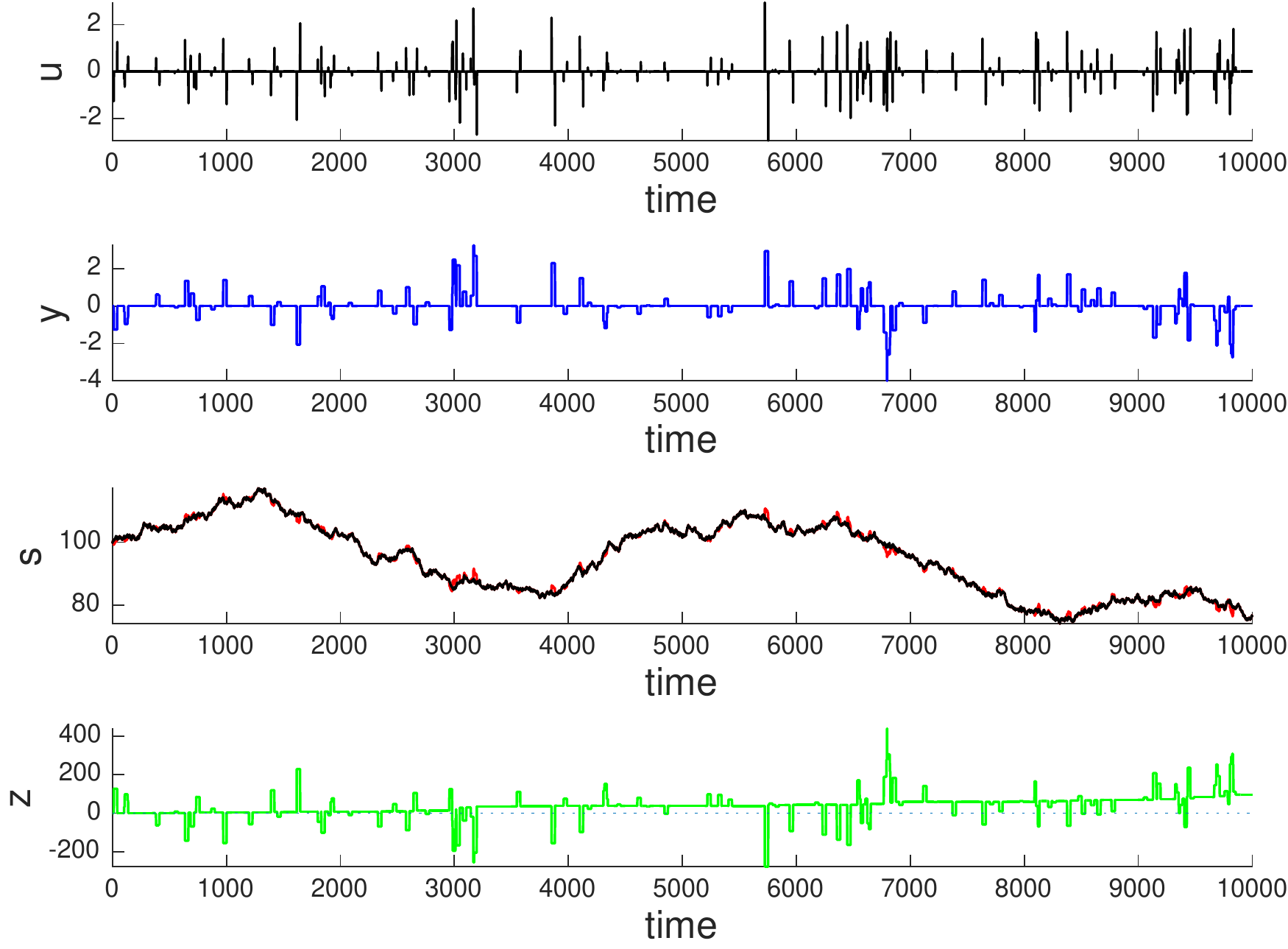}} 
\subfigure[]{
\includegraphics[width=5.5cm]{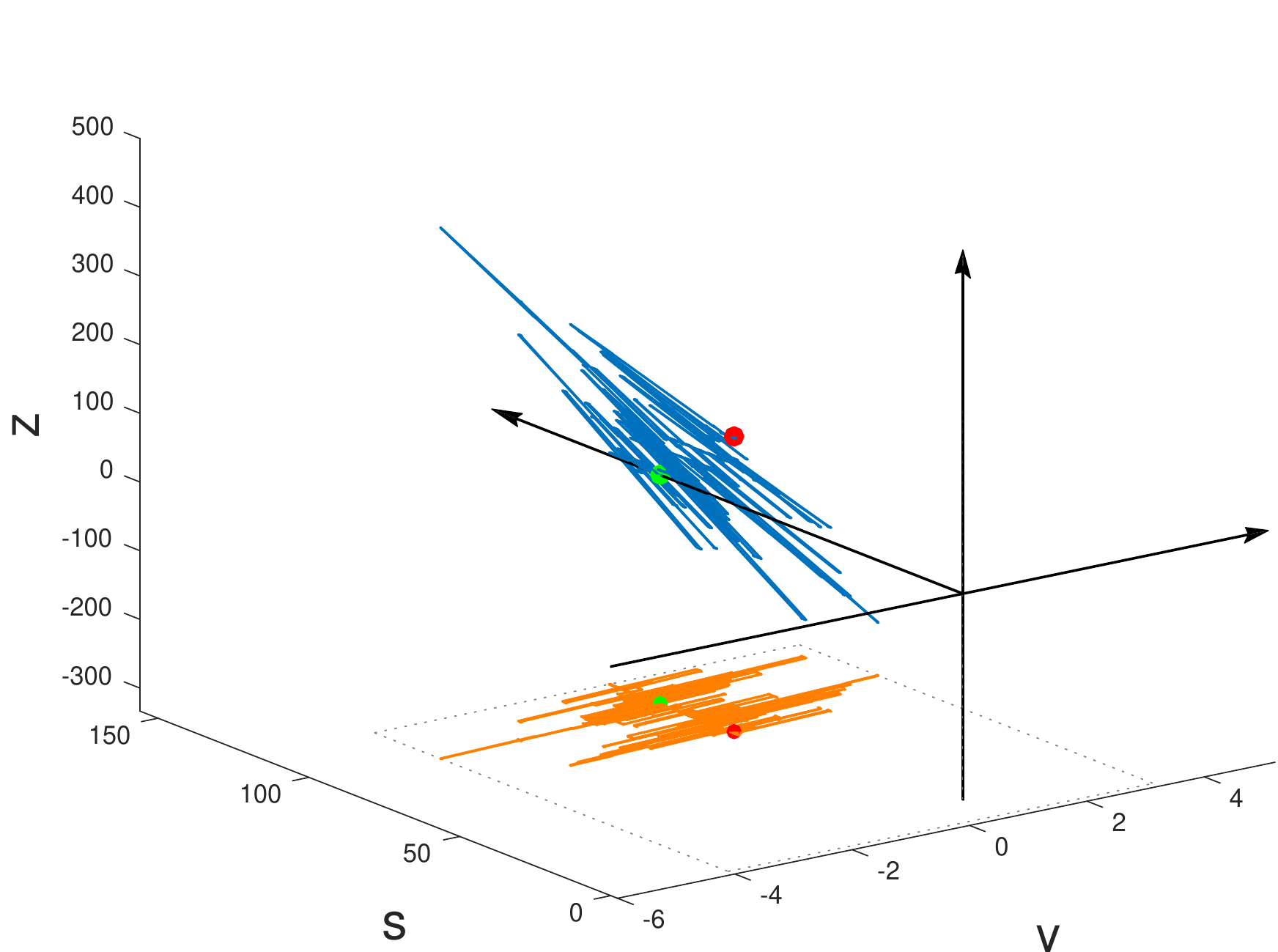}} 
\caption{Effect of cyclic trading operations. Panel (a): time profiles of the variables for a single buy-then-sell cycle. Panel (b): corresponding shape and phase space profiles. 
At time $ t_b=20$ the trader buys a unit of stock.
The stock quote $s$ increases consequently because of the price impact of the operation, while the cash balance $ z$ becomes negative.
At time $ t_s =80$ the stock is sold and the stock quote returns to the level it had at $t=0$. 
However, the cash balance $ z$ becomes positive because the stock is sold at a price higher than it was bought. 
The shape trajectory of the variables $ (y, \, s )$ is cyclic and has zero area (orange curve in panel (b)).
The phase variable $ z$ experiences a net displacement because of the cycle, i.e., a geometric phase is produced (blue curve in panel (b); the starting point is the green dot and the final point is the red dot).
Panels (c) and (d):  When the stock quote is allowed to drift because of the operations of other traders, then periodic operations of a trader no longer result in cyclic trajectories in shape space $ (y, \, s)$. However a geometric phase in the $ z$ variable is still present and keeps building up whenever the trader completes a cycle of operations.
In panel (c), the plot for $ s$ shows in black the stock quote in absence of trading cycles, and in red the deviations due to these trading cycles, visible only during the intervals in which stocks are hold (or shorted, for sell-then-buy cycles).}
\label{fig:cycle1}
\end{center}
\end{figure}

\begin{figure}[htb!]
\begin{center}
\subfigure[]{
\includegraphics[width=6cm]{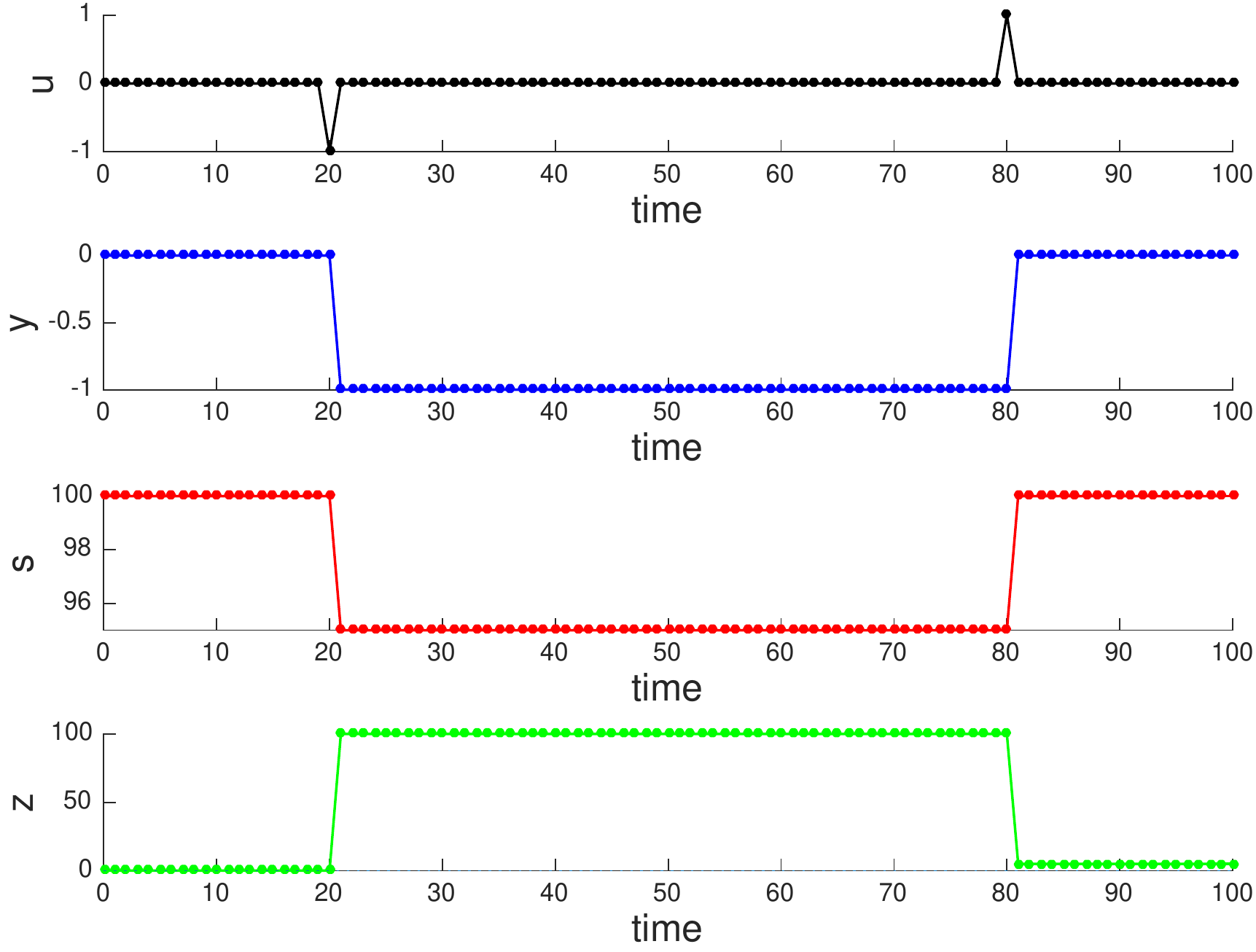}} 
\subfigure[]{
\includegraphics[width=6cm]{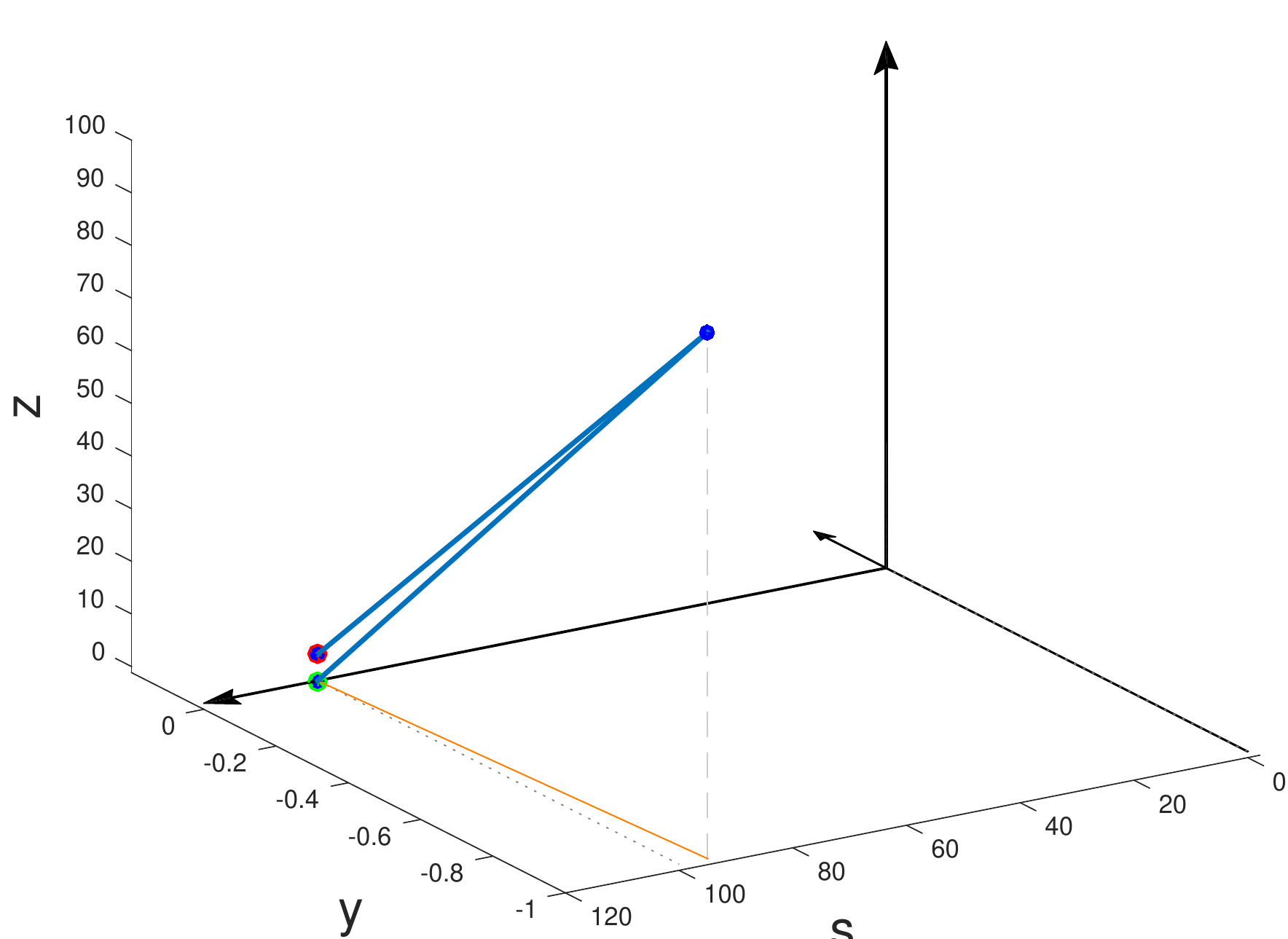}}
\caption{A sell-then-buy cycle. Panel (a): time profiles of the variables. Panel (b): shape and phase space profiles. 
Also in this case the trader obtains a profit from the cyclic operation (red dot in panel (b) is above the green dot), because the stock is bought at a lower price than it was sold.}
\label{fig:cycle2}
\end{center}
\end{figure}

Assume that the trader performs a cyclic operation, consisting in buying a certain amount $ k$ of stocks at time $ t_b $ and then reselling them at a later time $ t_s $. 
The time unfolding of events for this ``buy-then-sell'' cycle is shown in Fig.~\ref{fig:cycle1}, panels (a) and (b). 
Denote $ t=0 $ the starting time, and $ t_\text{end} $ any time larger that $ t_s$. 
The buy-then-sell cycle corresponds to the input pattern (see Fig.~\ref{fig:cycle1}(a), top subplot)
\beq
u(t) = \begin{cases} 
0 & t\in [ 0, \, t_{b-1} ] \\ 
k & t = t_b \\
0 & t \in [ t_{b+1} , \, t_{s-1}  ] \\ 
-k & t = t_s  \\
0 & t \in [ t_{s+1}, \, t_\text{end} ] .
\end{cases}
\label{eq:input_cycle}
\eeq
If no other trading operation happens during the time interval, the evolution of the state vector is
\beq
x(t) = \begin{cases} 
x(0) & t \in [0, \, t_b ] \\ 
x(0) + \begin{bmatrix} k \\ r k  \\ - s(0) k \end{bmatrix} & t \in [ t_{b+1}, \, t_s ] \\
x(0) + \begin{bmatrix} 0 \\ 0  \\ r k^2 \end{bmatrix} & t \in [ t_{s+1}, \, t_\text{end} ] .
\end{cases}
\label{eq:state_cycle}
\eeq
In other words, the net amount of stock at the end of the cycle is equal to that at the start: $ y(t_\text{end} ) = y(0) $. 
When no other operation happens during the time interval then also $ s(t_\text{end} ) = s(0) $.
However, if $ r>0$, the cash balance is positive, i.e, $ z(t_\text{end}) = z(0) + r k^2 > z(0) $.
Such a positive cash balance is solely due to the price impact of the transactions being made by our trader. 
It is proportional to the square of the number of stocks traded. 
A similar result occurs for a sell-then-buy cycle, which also leads to $ z(t_\text{end} ) > z(0)$, see Fig.~\ref{fig:cycle2}.
In both cases, while $ y(t) $ and $s(t)$ accomplishes a cyclic trajectory returning to their starting point, $ z(t)$ does not.
The key in understanding why a nonzero cash flow is generated by cyclic operations lies in the nonlinearity of the difference equation \eqref{eq:basicmodel3}, namely $ s(t) u(t)$ is a bilinear term, i.e., linear, but simultaneously so, in both state and input.
This nonlinearity introduces a behavior which is typical of nonintegrable vector fields \cite{Bloch4,murray1994mathematical}. 

The classical description of geometric phase deals with nonlinear continuous-time vector fields, in which 2 or more of the variables (corresponding to our $ y(t) $ and $ s(t)$) evolve on a shape space, and the remaining variables (here $ z(t) $) on a phase space above the shape space, see Fig.~\ref{fig:geometric_phase}(a).
Cyclic trajectories in shape space yield a net motion on the phase variable which is proportional to the area of the cyclic shape trajectory and it is zero when this area is zero, see Fig.~S5. 

In discrete-time, however, the picture is different. 
As shown in Fig.~\ref{fig:cycle1} and as computed explicitly in \eqref{eq:state_cycle}, even a zero-area cyclic trajectory in shape space yields a non-zero motion on the phase variable. 
In our specific example, the origin of the positive cash flow induced by a cyclic trajectory (entering then exiting a position) is due to the fact that stocks are bought at a lower price $ s(t_b)$ and are sold at a higher price $ s(t_s)$, see Fig.~\ref{fig:cycle1}. 
The price increase is due to our assumption that stock trades impact stock quotes. 
The argument is analogous if the cycle consists of a sell-then-buy: in this case the quote at selling time is higher than that at buying time, see Fig.~\ref{fig:cycle2}.

\paragraph{Effect of a market drift.}
Let us now consider a scenario in which the stock price is driven by the operations of our trader but also by a random variable $ \{ w(t_i) \} \in \mathcal{N}(0, \sigma) $ representing the trading operations of all other traders of the stock we are considering. 
The system \eqref{eq:basicmodel1}-\eqref{eq:basicmodel3} modifies as
\beqa
y (t_{i+1} ) & = &  u (t_i) \label{eq:stochasticmodel1} \\
s (t_{i+1} )  & = & s(t_i) + w(t_i) + r u (t_i)  \label{eq:stochasticmodel2} \\
z (t_{i+1})  & = &z(t_i) - s (t_i)  u (t_i) \label{eq:stochasticmodel3} 
\eeqa
where we still assume that operations of our trader are cyclic and follow \eqref{eq:input_cycle}, but eq. \eqref{eq:stochasticmodel2} has an extra drift term.
At the end of a buy-then-sell cycle, instead of \eqref{eq:state_cycle} we have:
\beq
x(t_\text{end}) = \begin{bmatrix} 
y(0) \\ s(0) + \sum_{i=0}^\text{end} w(t_i) \\
z(0) +  \sum_{i=b}^s w(t_i) k + r k^2 \end{bmatrix}
\label{eq:stochasticmodel1_sol}
\eeq
i.e., the stock trend in the hold interval $ [t_b, \, t_s ] $ also affects the cash flow of our trader, but only as a first order term of the number of stocks held during the cycle, as opposed to the second order term of the noncommutative effect. 
Iterating over many cyclic operations of our trader, initiated at time instants $ \{ t_b \} $ drawn from a Poissson process, and drawing also $ \{ t_s - t_b  \} $ from a Poisson distribution of shorter mean arrival time, then the stochastic difference equation \eqref{eq:stochasticmodel1}-\eqref{eq:stochasticmodel3} is such that at $ t_\text{end} $, $ y (t_\text{end}) = y(0)$, while instead now the stock quote has a nontrivial trend $ s (t_\text{end}) \neq s(0)$.
From \eqref{eq:stochasticmodel1_sol}, the influence of the stock trend on the cash balance can be made small by reducing the length of the time interval $ [ t_b, \, t_s ] $, so that regardless of the value of the drift term $ \{ w (t_i) \} $, it is possible to obtain $ z(t_\text{end}) > z(0) =0 $.
Hence, provided that a cycle is completed at sufficiently high frequency, the price impact induced by the cycle itself can still dominate the trend imposed by the market.
A sample trajectory depicting this situation is shown in Fig.~\ref{fig:cycle1} (c) and (d).
In panel (c), in the plot of $ s$, the stock quote driven only by $ \{ w(t_i) \} $ is shown in black. 
The red deviations appearing in the plot correspond to the price impact of the operations carried out by our trader. 
In fact, combining \eqref{eq:state_cycle} and \eqref{eq:stochasticmodel1_sol} for a single buy-then-sell cycle the value of $ s $ in the holding interval $ [t_b, \, t_s ] $ is given by 
\[
s(t_i) = s(0) + \sum_{j=b}^{i-1} w(t_j) k + r k, \qquad t_i \in [t_{b+1}, \, t_s ]
\]
i..e, the extra term $ rk $ must be added to $ \{ w(t_i) \}$. 

\begin{figure}[htb!]
\begin{center}
\includegraphics[angle=0, trim=2cm 5cm 2cm 5cm, clip=true,width=10cm]{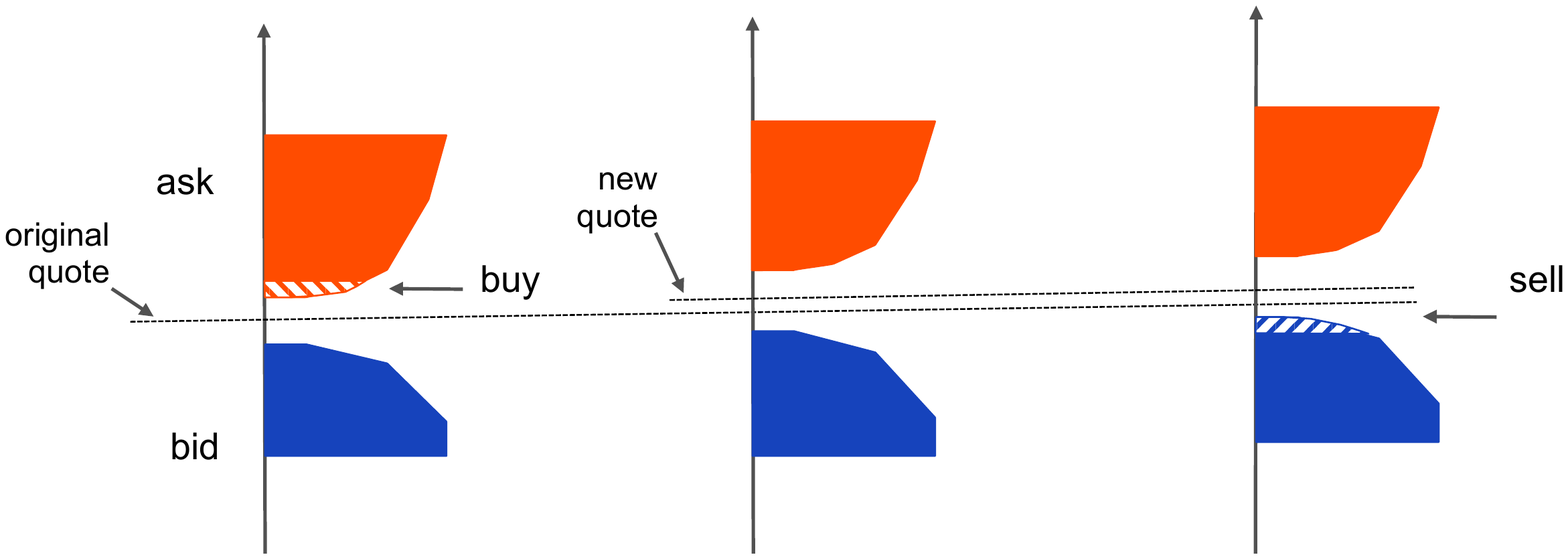}
\caption[]{Effect of a wide bid-ask spread. The buy order empties the shaded red region of the ask side of the limit order book. 
The stock quote (midprice) consequently increases, while the best bid price (blue) remains unchanged. The sell order needed to complete the cycle settles at a price below the original buy price, hence the cycle results in a net loss for the trader.}
\label{fig:ask-bid-spread1}
\end{center}
\end{figure}

\begin{figure}[htb!]
\begin{center}
\includegraphics[angle=0, trim=2cm 5cm 2cm 5cm, clip=true,width=10cm]{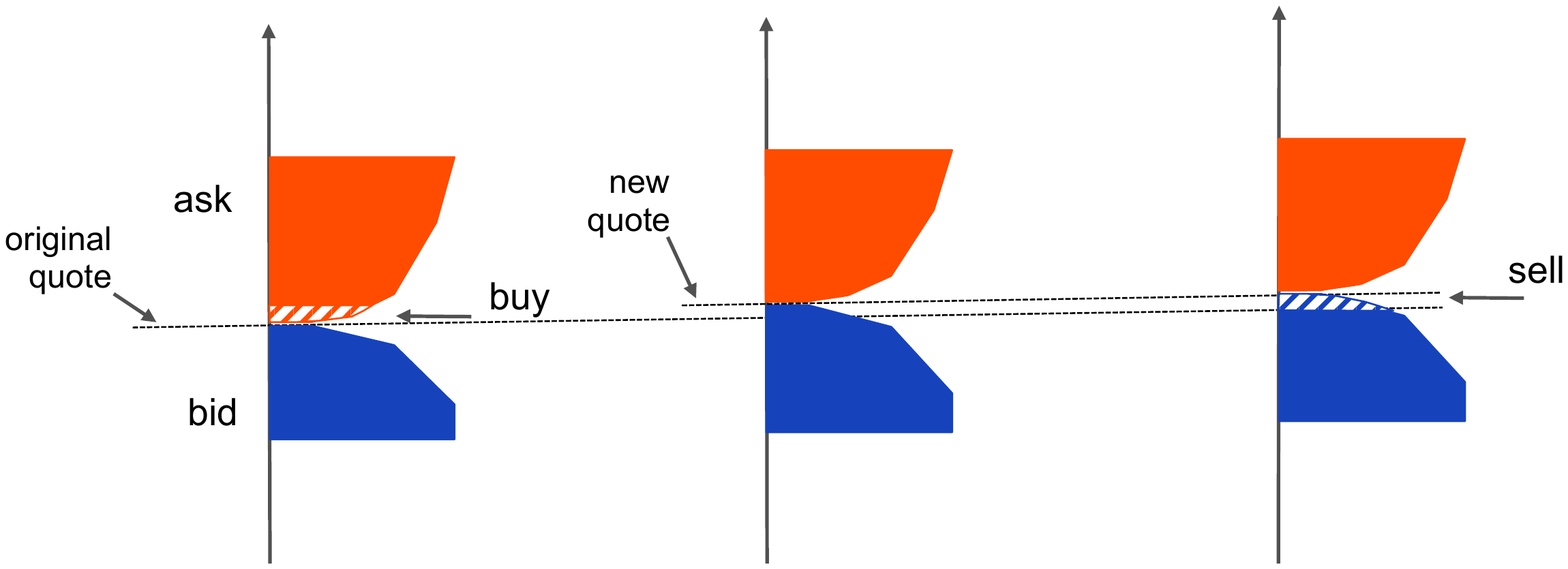}
\caption[]{Effect of a small bid-ask spread. In a deep market with many buy and sell orders arriving continuously, the spread remains small around the midprice, even after the execution of the buy order has raised the stock quote.
Hence the price impact of the buy order can lead to a new best bid price which is above the quote at which the buy order was executed. In this case, completing the cycle indeed results in a net  gain. }
\label{fig:ask-bid-spread2}
\end{center}
\end{figure}

\paragraph{Effect of bid-ask spread.}
In real stock exchanges a buy (market) order is compared with a limit order book: the limit order with best ask price is matched to the market order, and if needed the book is ``walked'' to the next ask offer and so on until the order is completely executed (order redirection towards other stock exchanges is neglected here) \cite{Gould2013Limit,cartea2015algorithmic}. 
Consequently the spread between the best ask and best bid price increases and the midprice (which in our model is the stock quote) is pushed up. 
If nothing else happens after the execution of the order, then when trying to complete the cycle the best bid price is unchanged (and below both the new and old original stock quote), hence the cycle results in a net loss rather than a profit as we are showing here, see Fig.~\ref{fig:ask-bid-spread1}.
However if the market has sufficient depth and the bid-ask spread is small and remains small throughout the process, for instance because new limit orders on both ask and bid side keep arriving (``locked'' or ``nearly locked'' markets), then it may still happen that a raising stock price drives the new best bid price above the old ask price (value at which the stock was bought), see Fig.~\ref{fig:ask-bid-spread2}.

\begin{figure}[htb!]
\begin{center}
\subfigure[]{
\includegraphics[width=6cm]{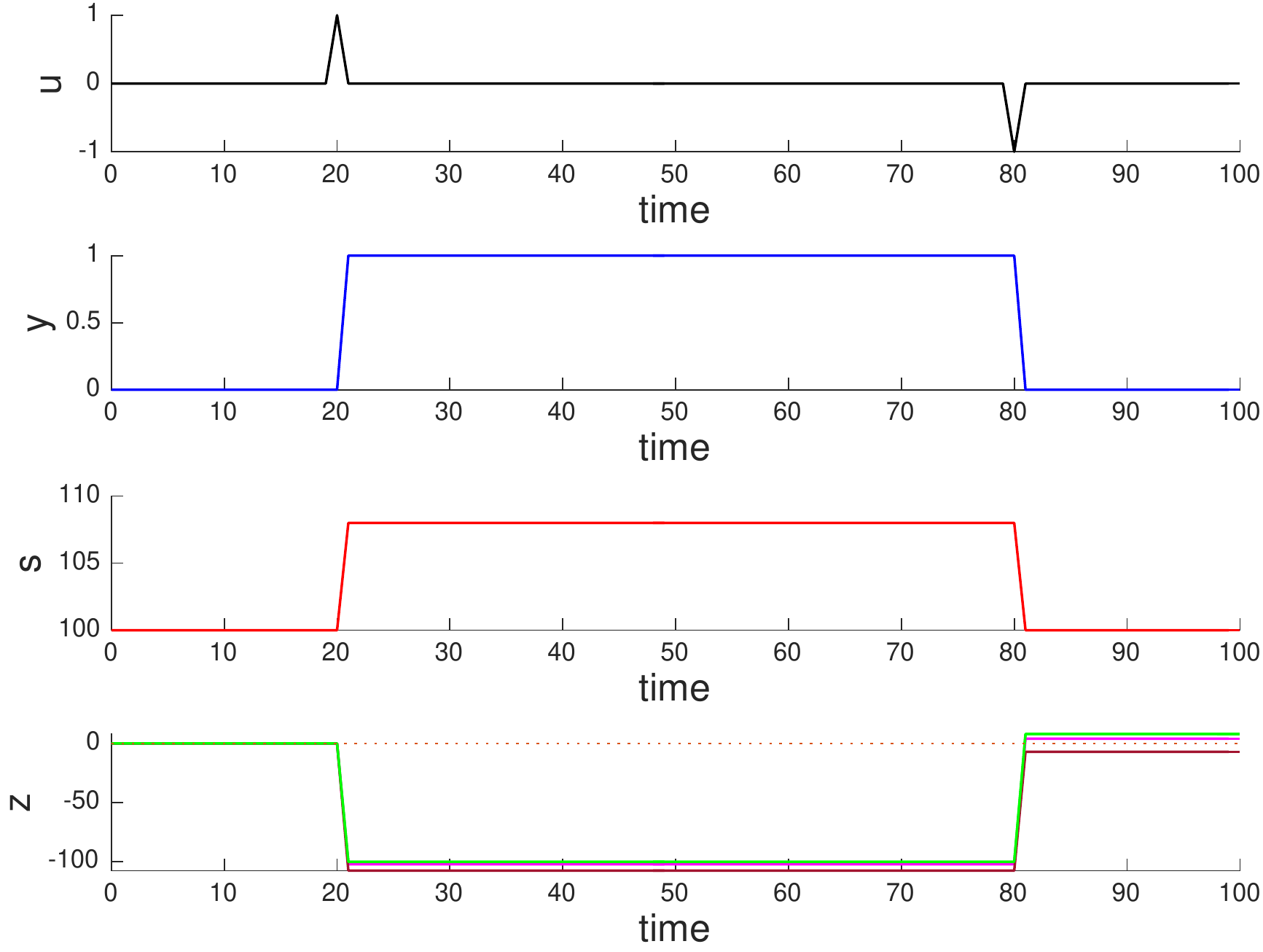}} $ \qquad $ 
\subfigure[]{
\includegraphics[width=6cm]{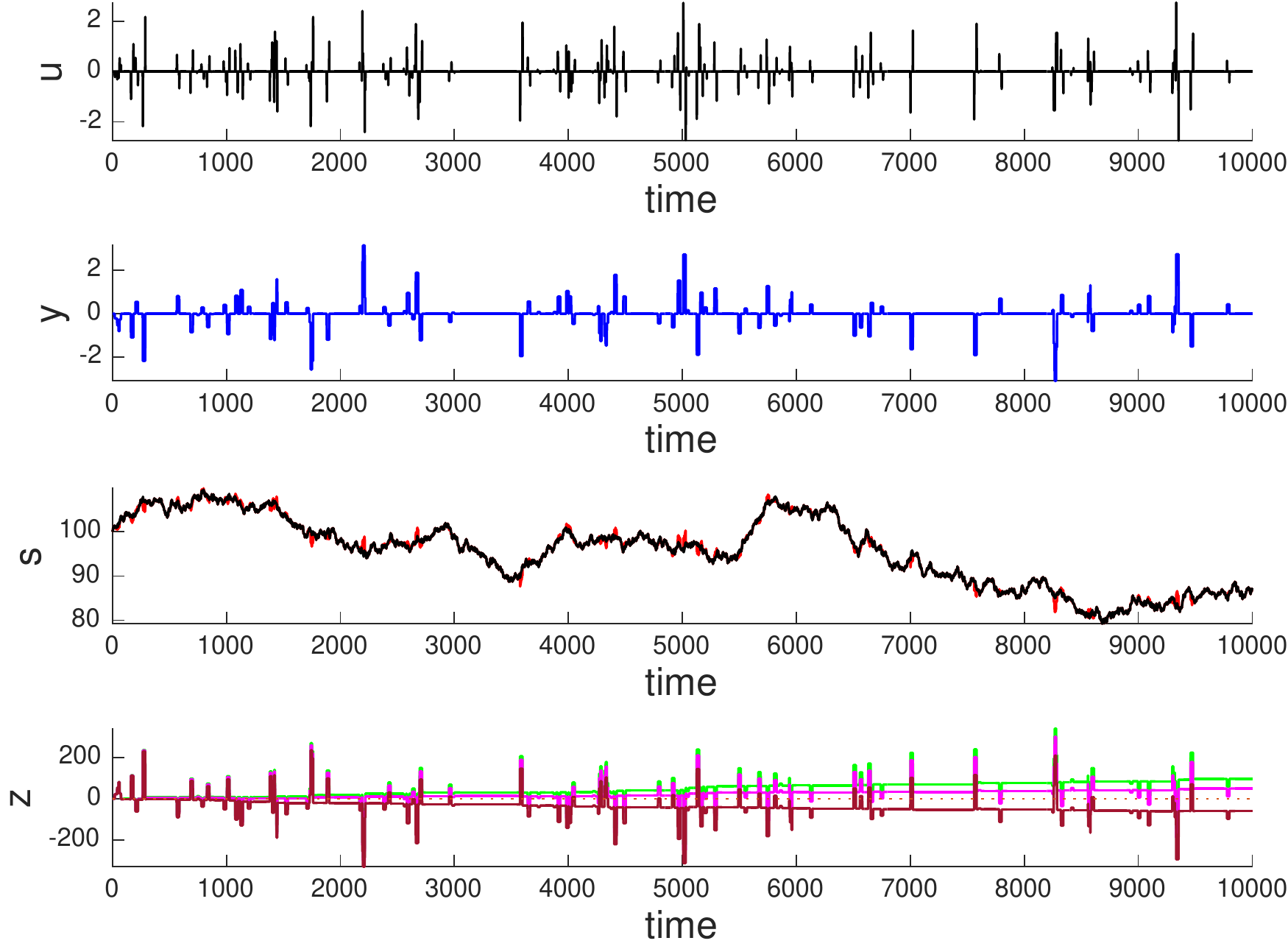}}
\caption{Spread reduces profits (or induces losses).  Panel (a): In \eqref{eq:sol:spread}, the presence of a quote spread decreases the value of $ z(t_\text{end}) - z(0) $. In the plot for $ z$, if the green curve corresponds to $ q=c=0$, the magenta and brown curves correspond to different choices of these parameters. Depending on the values of $ r$, $q$ and $ c$, the geometric phase can correspond to a profit (green and magenta curves), or to a loss (brown). The geometric phase (and the corresponding profits or losses) accumulates over repeated  trading cycles (panel (b)).
For short trading cycles the presence of a stock drift (panel (b)) does not alter the result.} 
\label{fig:spread}
\end{center}
\end{figure}

Assume that the stock quote $ s(t_i)$ is equal to the midprice between best ask $ a(t_i) $ and best bid $b(t_i)$ prices, and that the bid-ask spread $ q$ around $ s(t_i) $ is constant (i.e., $ a(t_i) = s(t_i) + q/2 $, and $ b(t_i) = s(t_i) - q/2 $).
To take into account the bid-ask spread in our model, eq. \eqref{eq:basicmodel3} must be replaced by 
\beq
z(t_{i+1})  =  z(t_i) - v(t_i) u(t_i) \qquad  \text{where  } v(t_i) = 
\begin{cases} 
a(t_i) & \text{ if } u(t_i) >0 \\
b(t_i) &  \text{ if } u(t_i) <0 .
\end{cases}
\label{eq:basicmodel3spread} 
\eeq
At the end of the cycle, the cash balance from \eqref{eq:basicmodel3spread} is $ z(t_\text{end}) = z(0) + r k^2 - q k $. 
If we include also a fee for each trading operation (assumed constant and equal to $c$), then 
\beq
z(t_\text{end}) = z(0) +  r k^2  - q k - 2 c .
\label{eq:sol:spread}
\eeq
Hence a cycle yields a net profit if $ z(t_\text{end}) - z(0) = r k^2 - qk - 2c >0$. 
See Fig.~\ref{fig:spread} 
for an example of how spread can reduce gains (or induce losses) in $ z(t)$.
In \eqref{eq:sol:spread}, if the market is deep enough and the spread limited enough, then $ z(t_\text{end} ) - z(0) $ can be made positive by choosing a sufficiently high value for $ k$.
Regardless of its sign, the quantity $ z(t_\text{end} ) - z(0)$ is still a geometric phase, representing profit when $ z(t_\text{end} ) - z(0) > 0$ or losses when $ z(t_\text{end} ) - z(0) < 0$.

\begin{figure}[h]
\begin{center}
\subfigure[]{
\includegraphics[width=5.5cm]{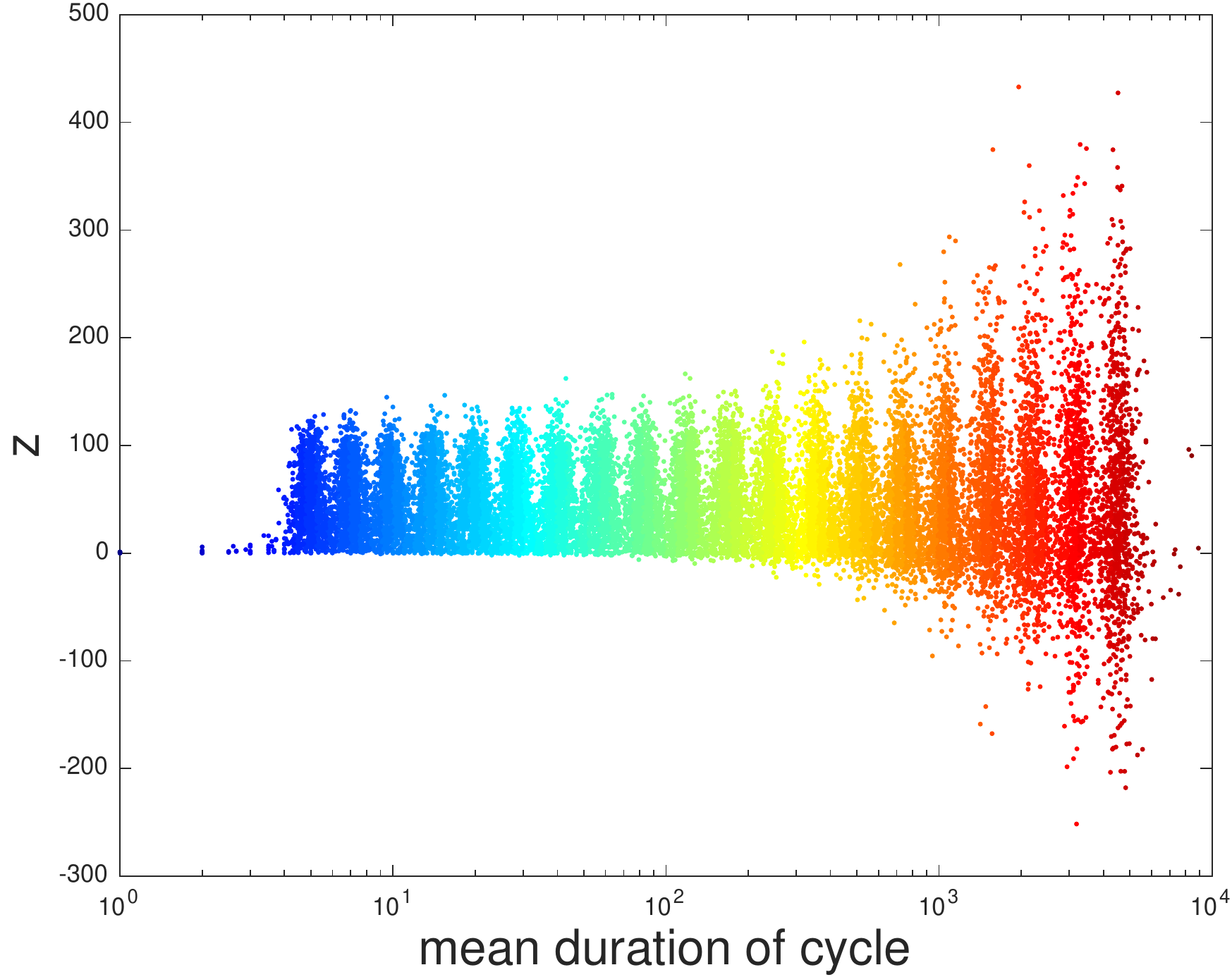}} $ \qquad $ 
\subfigure[]{
\includegraphics[width=5.5cm]{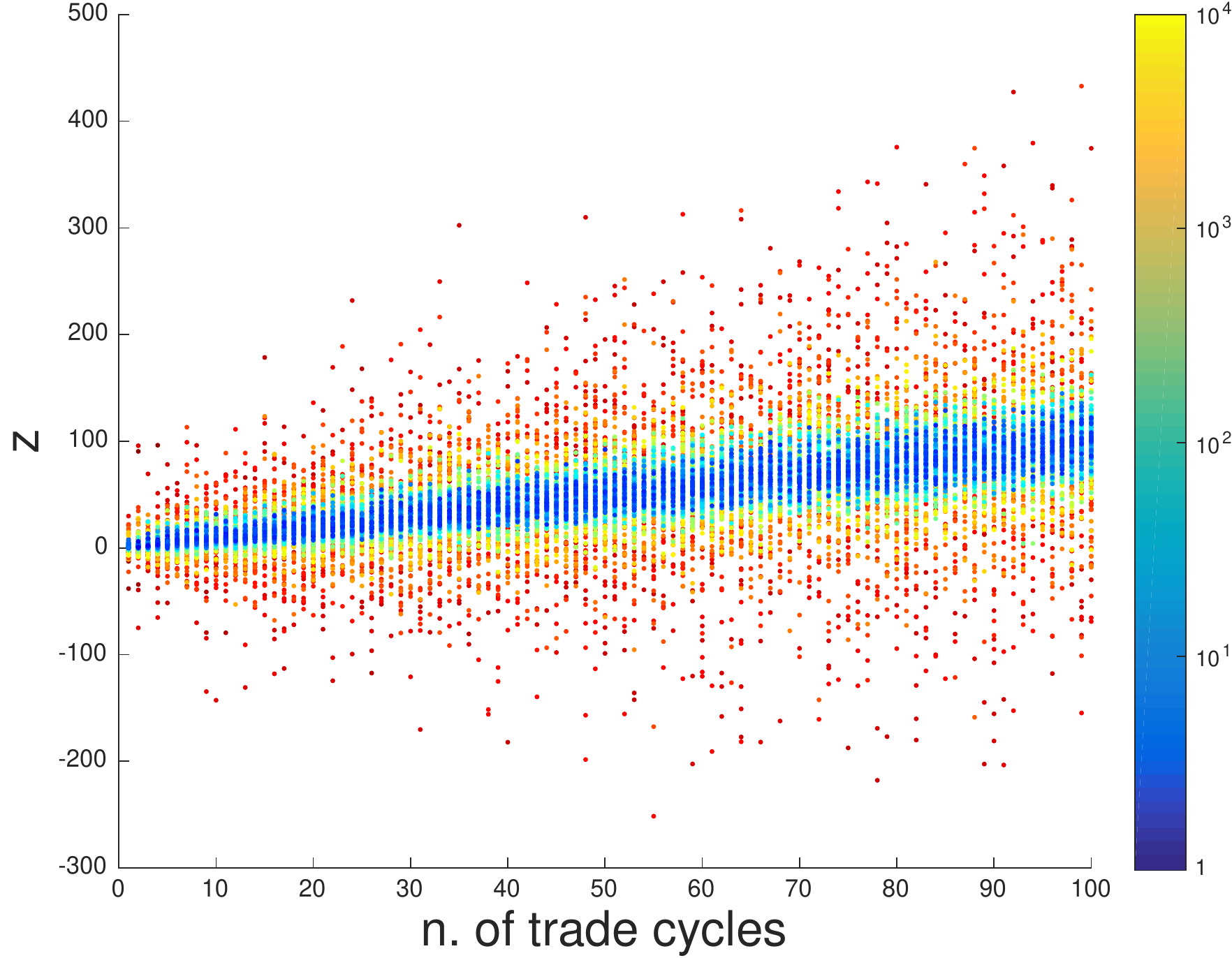}} \\
\subfigure[]{
\includegraphics[width=5.5cm]{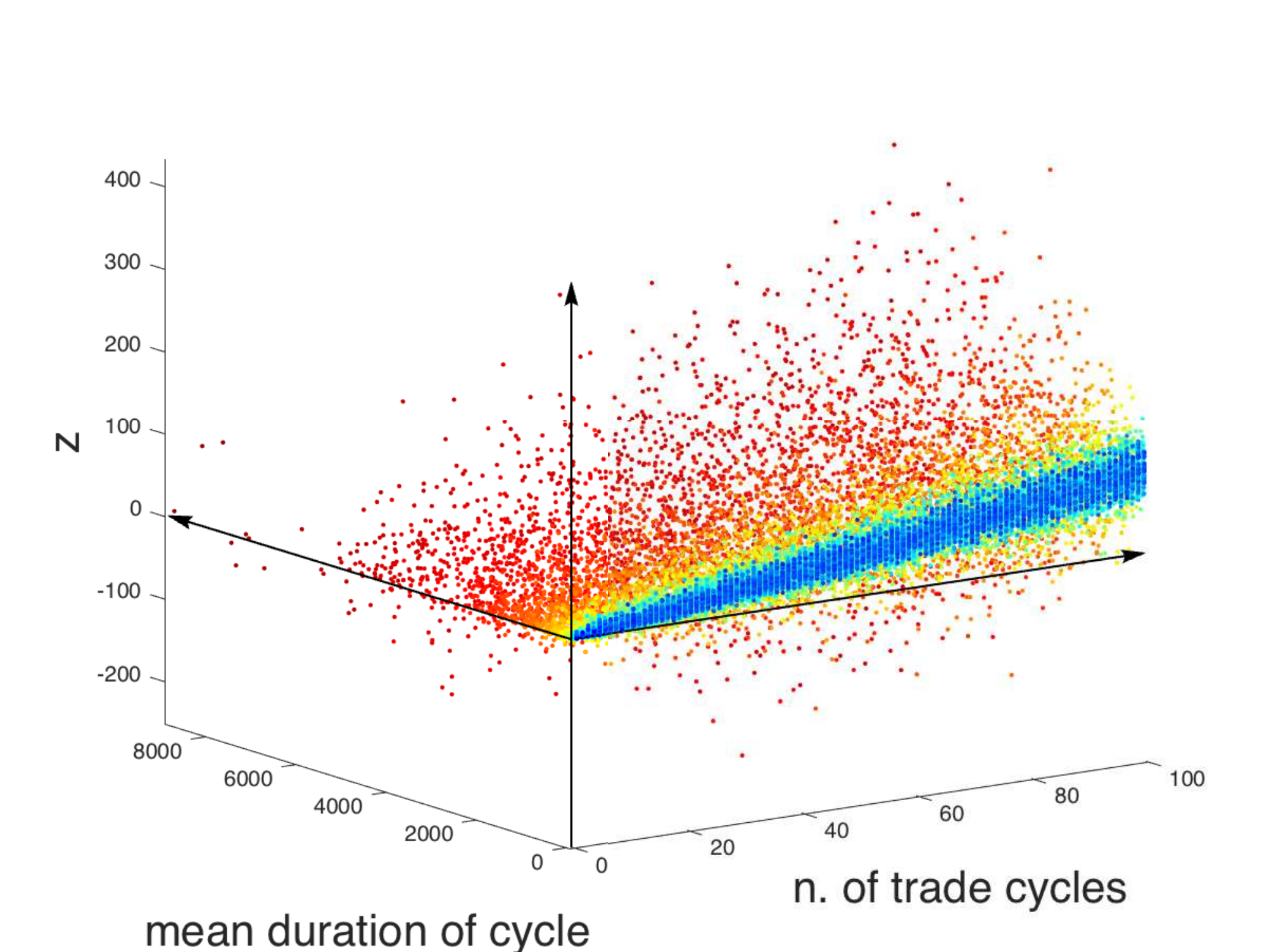}} $ \qquad $ 
\subfigure[]{
\includegraphics[width=5.5cm]{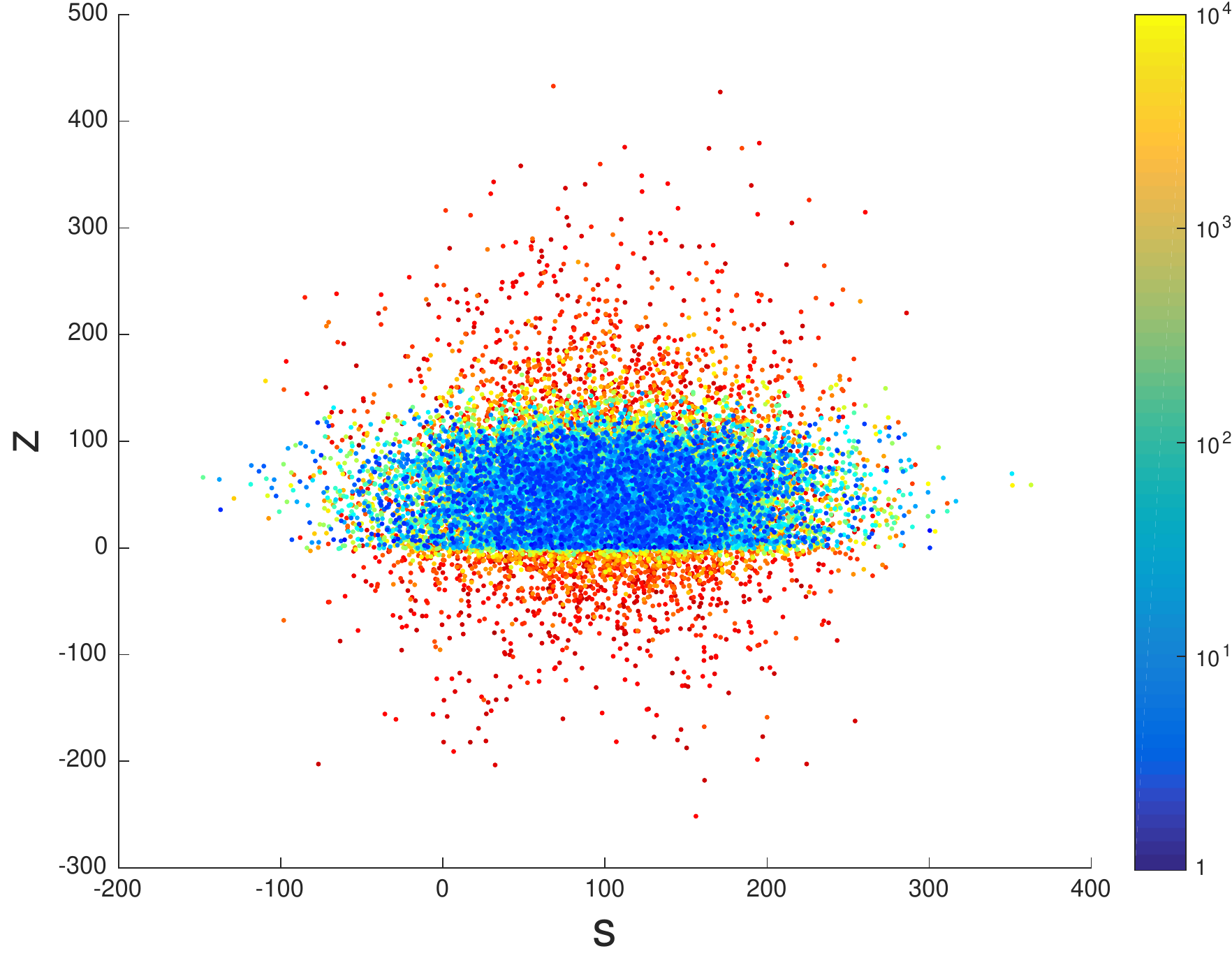}} \\
\caption[]{Endpoint of a series of cyclic trading operations, in presence of stock quote spread and quote drift. Each dot is the result of 1000 simulations, under the assumption that $ z(t_\text{end}) - z(0) \geq 0$ when the stock drift is absent (formula \eqref{eq:sol:spread}). The number of cyclic operations is considered itself a random variable (denoted ``n. of trade cycles''), and so is the duration of the holding intervals $ [t_b, \, t_s]$ (denoted ``mean duration of cycle''). 
For short cycles (with respect to the standard deviation $ \sigma $ of the stock drift term), the trader always obtains a profit (panel (a)). 
The profit is proportional to the number of trade cycles performed by the trader (panel (b)). The correlation tends to get lost when the mean duration of the trade cycles becomes large (red dots in panels (b) and (c)). For short cycles, the profit happens independently of the stock trend (blue dots in panel (d)). Also when the trade cycles become long and profit may be lost, there is no significant correlation between $ z$ and $ s$ (panel (d)).}
\label{fig:manycycle1_stat2}
\end{center}
\end{figure}

In  presence of a drift term $ \{ w(t_i) \} $ for the stock price as in \eqref{eq:stochasticmodel1}-\eqref{eq:stochasticmodel3}, the cash balance can be modified according to \eqref{eq:stochasticmodel1_sol} as
\[
z(t_\text{end} ) = z(0) + r k^2 + \left( \sum_{i=b}^s w(t_i) - q \right) k - 2c .
\]
As already mentioned, when the hold interval $ [t_b, \, t_s ]$ is short, so is the effect of the drift term 
$ \sum_{i=b}^s w(t_i) k $ on the final value of the cash balance. 
Fig.~\ref{fig:manycycle1_stat2} shows the value of $ z(t_\text{end})$ as we vary the number of trading cycles and the average duration of the hold intervals for a large number of realizations $ \{ w(t_i)\}$.
When in a formula like \eqref{eq:sol:spread} the geometric phase is positive ($ z(t_\text{end}) > z(0)$), then even in presence of drift the cash flow remains positive at least as long as $ [t_b, \, t_s ]$ is sufficiently short compared to $ \sigma$. 
Short-lived deviations in the stock quote are enough to manifest the noncommutativity effect we are describing here. 
In Fig.~\ref{fig:manycycle1_stat2}, notice how rather than to the quote drift, $ z(t_\text{end}) $ is correlated to the number of trading cycles carried out in $ [0, \, t_\text{end}]$ (which follows from $ r $ being a constant).

\paragraph{Effect of more complex price impact formulas.}
All models discussed in the paper make use of a linear and constant formula for describing the price impact of trading operations on the stock quote. 
However, any nonlinear function of the trader's orders can be used instead of \eqref{eq:basicmodel2} , for instance
\beq
s(t_{i+1}) = s(t_i) + r(u(t_i)) 
\label{eq:price_impact1}
\eeq
where $ r(u)$ is any time-invariant function of $ u$ such that 
\beq
\begin{cases} 
r(u) >0 & \text{if  } u >0 \\ 
r(u) <0 & \text{if  } u <0 .
\end{cases}
\label{eq:price_impact2}
\eeq
If $ r(\cdot ) $ is an odd function, i.e., $ r(-u) = - r(u) $, then the neutrality of the stock quote to cyclic trades is preserved, that is, in absence of drift $ s(t_\text{end} ) = s(0) $ and our arguments are still valid. 
For instance, combining \eqref{eq:price_impact1}-\eqref{eq:price_impact2} with the spread model \eqref{eq:basicmodel3spread}  and the constant trading fee $ c$ yields
\[
z(t_\text{end} ) = z(0) +  \left( r(k) - q \right) k - 2c 
\]
which becomes
\[
z(t_\text{end} ) = z(0) + \left( \sum_{i=b}^s w(t_i) + r(k) - q \right) k - 2c 
\]
when also the drift term $ \{ w(t_i) \} $ is considered.

\begin{figure}[htb!]
\begin{center}
\subfigure[]{
\includegraphics[width=6cm]{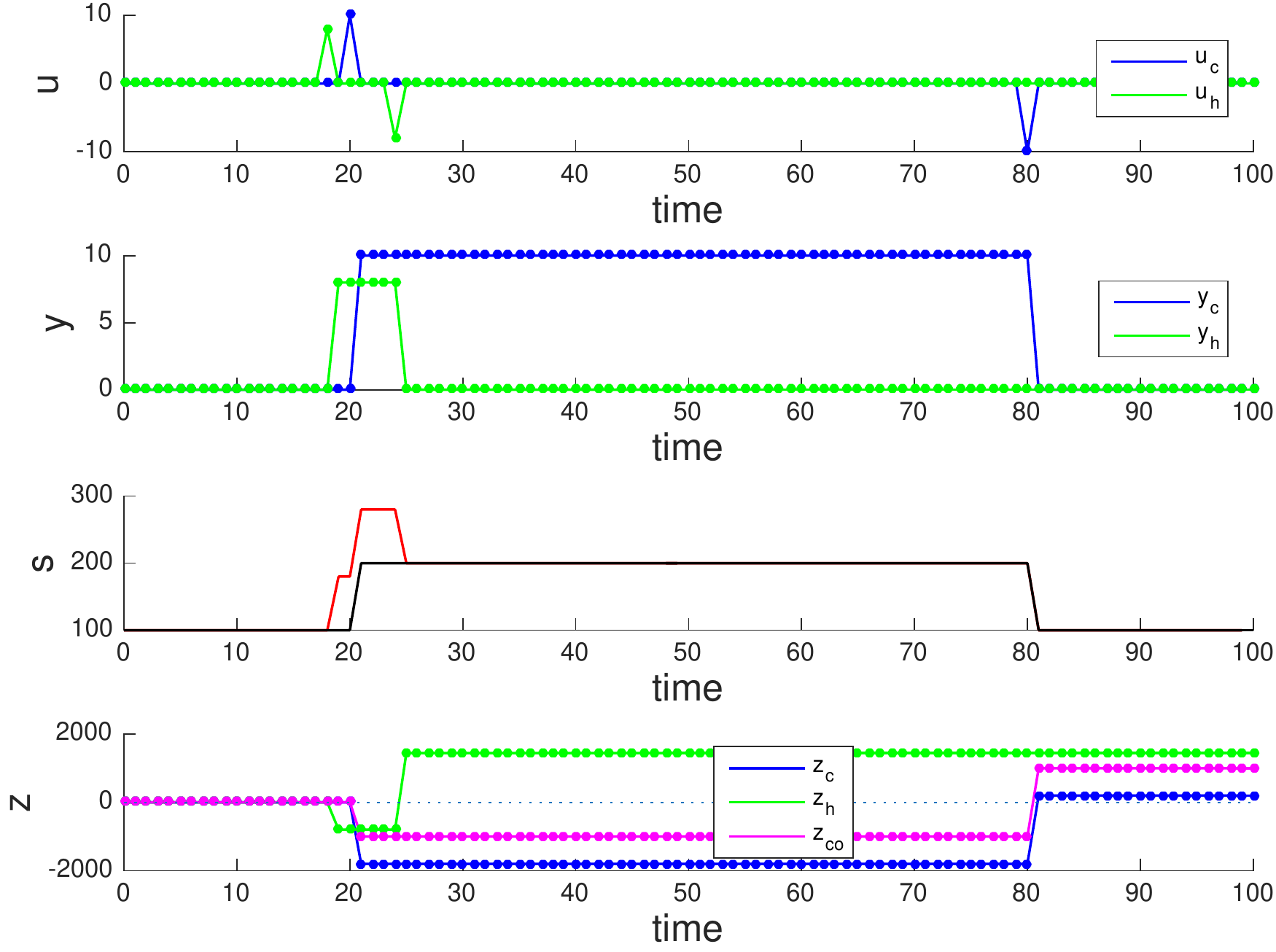}} $ \qquad $ 
\subfigure[]{
\includegraphics[width=6cm]{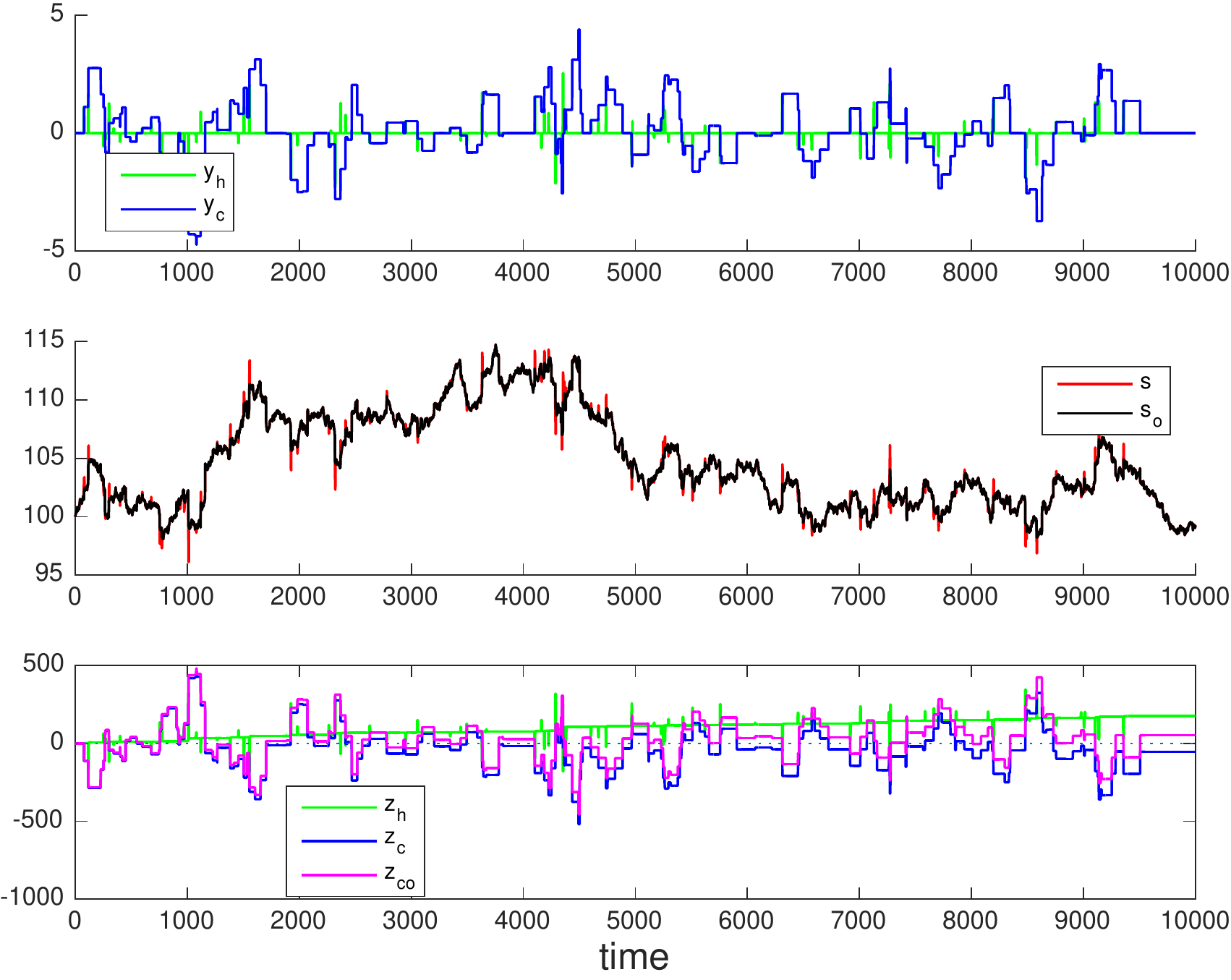}} 
\caption[]{Front-running an order. A high-frequency trader that can anticipate the operations of a slower classical trader can benefit of the price impact created by the latter, in addition to the one induced by his own trades.
The geometric phase (and hence any possible profit) tends to swap from the classical trader to the high-frequency trader.
For the classical trader, the profit in presence of a front-runner is reduced (blue curve in the lower plot of both panels) with respect to the same curve in absence of front-run (magenta curve). }
\label{fig:frontrun1}
\end{center}
\end{figure}

\paragraph{Gaining phase by front-running an order.}
When multiple traders are considered simultaneously, an ``easy'' way to guarantee that a trader has a positive cash balance is to have him front-run somebody's else orders, a strategy sometimes imputed to certain aggressive high-frequency traders \cite{lewis2014flash}. 
Consider two traders, one ``classical'' (indexed by $ c$) and one high-frequency (indexed by $ h$).
Assume that each order entered by the classical trader is executed slowly and front-run by the high-frequency trader.
Practically, this can be achieved by a combination of colocalization and order rerouting among different physical stock exchanges \cite{lewis2014flash}. 
If $ t_{c_b} $ and $ t_{c_s} $ are the execution times of the buy and sell order of the classical trader, then let $ t_{h_b} = t_{c_b} - \tau_1 $ and $ t_{h_s} = t_{c_b}+ \tau_2 $, $ \tau_1, \, \tau_2 >0 $, be those of the high-frequency trader. 
As shown in Fig.~\ref{fig:frontrun1}, a buy order initiated by the classical trader triggers a first price increase at $ t_{h_b} $ and a second one at $ t_{c_b}$. 
If the high-frequency trader resells his stocks right after $ t_{c_b}$, then he can benefit (also) of this second price increase to make a profit. 
Using equations similar to \eqref{eq:basicmodel1}--\eqref{eq:basicmodel3} to describe the system:
\beqa
y_\alpha(t_{i+1}) & = & y_\alpha(t_i) + u_\alpha(t_i) , \qquad  \alpha=\{ c, \, h \} 
\label{eq:ch_model1} \\
s(t_{i+1}) & = & s(t_i) + \sum_{\alpha=\{ c, \, h \} }  r u_\alpha (t_i) \label{eq:ch_model2} \\
z_\alpha(t_{i+1}) & = & z_\alpha (t_i) - s(t_i) u_\alpha (t_i)  , \qquad  \alpha=\{ c, \, h \}  .\label{eq:ch_model3} 
\eeqa
At the end of a cycle for both traders ($ t_\text{end} \geq t_{c_s} $), we have 
\beqa
y_\alpha (t_\text{end} ) & = & y_\alpha (0) , \qquad  \alpha=\{ c, \, h \} \nonumber \\
s(t_\text{end} ) & = & s(0)   \nonumber \\
z_c ( t_\text{end} ) & = & z_c (0) + r k_c (k_c-k_h ) \label{eq:ch_model1_spread_c}  \\
z_h ( t_\text{end} ) & = & z_h (0) + r k_h (k_c+ k_h ) \label{eq:ch_model1_spread_h} 
\eeqa
where $ k_c $ and $ k_h $ are the amount of stocks traded by the classical and high-frequency trader during the cycle. 
From \eqref{eq:ch_model1_spread_c}-\eqref{eq:ch_model1_spread_h}, it can be seen that for the classical trader  the ``benefit'' of the geometric phase in terms of cash balance is eroded when a high-frequency trader front-runs his orders.
In particular, if the quantities traded by the two actors are equal, $ k_h = k_c $, then
\[
\begin{split} 
z_c ( t_\text{end} ) & = z_c (0)  \\
z_h ( t_\text{end} ) & = z_h (0) + 2 r k_h ^2 
\end{split}
\]
i.e., the price increase of the classical trader in cyclic operations is completely lost, to the benefit of the high-frequency trader. 
If instead $ k_h > k_c $, then the cycle always results in a loss for the classical trader: $ z_c ( t_\text{end} ) < z_c (0) $.
Analogous considerations apply to the more complex models discussed in this paper, incorporating quote drift and quote spread.

\section{Discussion}
A continuous-time equivalent of the model  \eqref{eq:basicmodel1}--\eqref{eq:basicmodel3} is presented in the Appendix. 
For it, a periodic input trajectory induces a zero-area cyclic path in shape space, which however results in no neat motion of the phase variable, see Fig.~\ref{fig:cont1}. 
Only when the area of the cyclic path in shape space becomes nonzero a geometric phase can appear, but for that to happen we have to modify the continuous-time model (for instance adding a time delay in the update law for the stock quote, see Fig.~S6\ref{fig:cont2}).
This confirms that the geometric phase described in this paper is intrinsically a discrete-time phenomenon, with no continuous-time counterpart.

Repeating a multitude of cycles on a fast time scale is a hallmark of high-frequency trading, which nowadays in certain equity markets like the US stock market constitutes around 50\% of the total trading volume \cite{duhigg_2009,aldridge2013high,Carrion2013680}.
There is no consensus in the literature on what are the sources of profit of high frequency trading operations of this type \cite{Brogaard14052014,Carrion2013680,cartea2015algorithmic,Hasbrouck2013646,Menkveld2013712}, other than ``penny'' profits must accumulate from each trading cycle. 
What the existence of a geometric phase suggests is that in principle tiny profits (or losses, when $ z(t_\text{end} ) - z(0) <0$) can be made through pure speculation, without holding any inventory position except for imperceptible amounts of time. 

If information on future market operations is available (like when speed is used to front-run upcoming orders), then this can be used to increase the chances of making profits by incorporating the geometric phase induced by the operations of the trader being front-run into that of the front-runner doing a cycle. 

Another conceptual consideration that follows from our models is that, as long as one considers cyclic operations, in principle profits or losses can be made with stock trading in a way that does not affect the main observable of a stock market, namely the quote of a stock. 
Making such profits (or losses) does not require any knowledge of market fundamentals, as the geometric phase just uses the price momentum induced by the act of trading itself, a bit like a self-fulfilling prophecy.
And this notwithstanding any ``efficient market'' theory. 

\section*{Appendix}

\paragraph{Classical interpretation of geometric phase in continuous-time.}

The continuous-time equivalent of \eqref{eq:basicmodel1}--\eqref{eq:basicmodel3}  is the following system of ODEs:
\beqa
\dot y(t) & = &  u(t) \label{eq:continuous_model1} \\
\dot s(t) & = &  r u(t) \label{eq:continuous_model2} \\
\dot z(t) & = &  - s(t) u(t) \label{eq:continuous_model3} .
\eeqa

The geometric interpretation of the result shown in Fig.~\ref{fig:geometric_phase} (a) for the model \eqref{eq:continuous_model1}-\eqref{eq:continuous_model3} is that if we consider $ x\in \mathcal{M} \subset \mathbb{R}^3 $ and the projection to the shape space $ \mathcal{S}\subset \mathbb{R}^2 $
\[
\begin{split}
\pi \, : \, \mathcal{M} & \to  \mathcal{S}  \\
x & \mapsto  \begin{bmatrix} y \\ s \end{bmatrix}
\end{split}
\]
then, given $ x(0) $, for each trajectory $ \gamma \,: \, [0, t] \to \mathcal{S} $ $ \exists $ a unique $ x (t) \in \mathcal{M} $ such that for the solution of \eqref{eq:continuous_model1}-\eqref{eq:continuous_model3}, $ x(t) = \pi^{-1} \left( \pi (x(t) \right) $, i.e., the geometric phase variable $z (t) $ corresponding to $ \gamma (t) $ is unique. 
In particular, if $ \Gamma \,: \, [0, T] \to \mathcal{S} $ is a closed shape curve enclosing an area $ \Omega $, then the geometric phase (or ``holonomy'', \cite{Marsden2,Bloch4}) of $ \Gamma $ is 
\[
z(T) = z(0) + \oint_\Gamma s \, d y 
\]
or, by Stokes theorem, 
\beq
\begin{split}
z(T) & = z(0) + \int_\Omega d(s \, d y ) \\
&  = z(0) + \int_\Omega d s \, d y  = z(0) + \omega
\end{split}
\label{eq:stokes} 
\eeq
where $ \omega $ is the area of $ \Omega$.
It follows that when the area encircled by the cyclic trajectory $ \Gamma $ in shape space is zero (i.e., $ \omega =0$), then the phase variable $ z $ must shows no net displacement at the end of the cycle.

\begin{figure}[htb!]
\begin{center}
\subfigure[]{
\includegraphics[width=6cm]{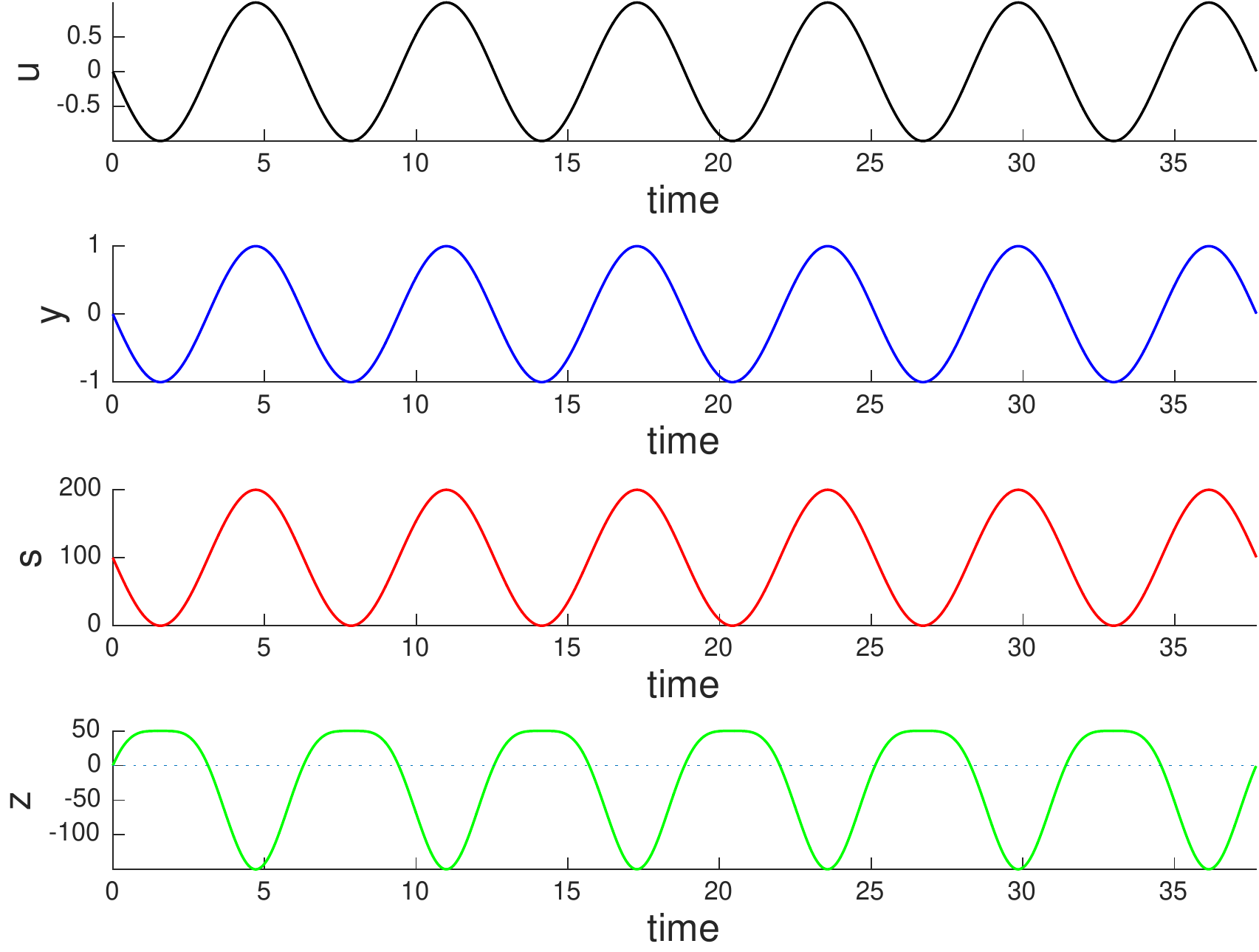}} $ \qquad $ 
\subfigure[]{
\includegraphics[width=6cm]{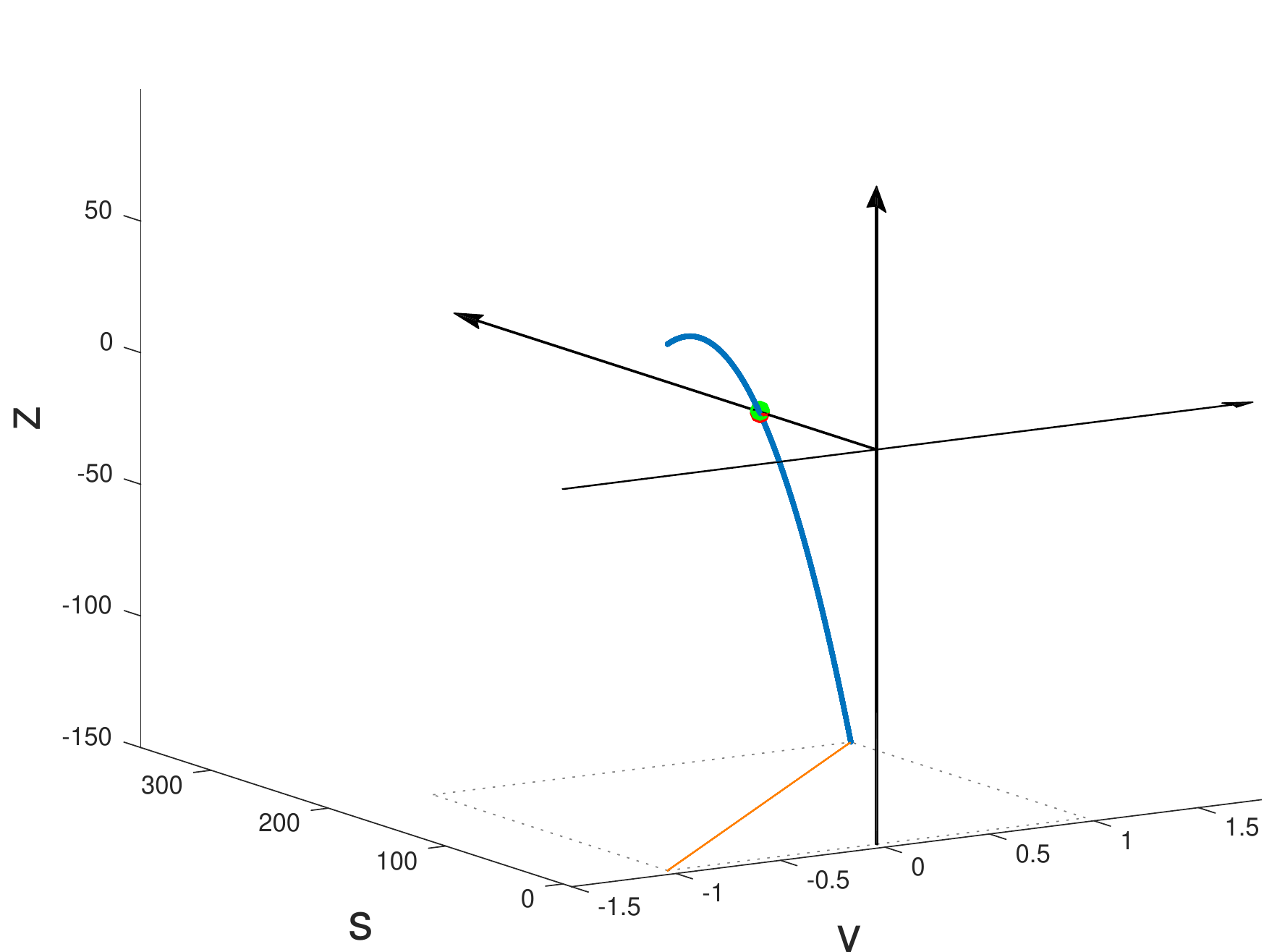}}
\caption{Continuous-time stock trading model: cycles of zero area in shape space (orange curve in panel (b)) yield no geometric phase.
The starting point (green dot in panel (b)) and the final point (red dot in panel (b)) overlap.}
\label{fig:cont1}
\end{center}
\end{figure}

Indeed in the system \eqref{eq:continuous_model1}-\eqref{eq:continuous_model3}, a cyclic trajectory in $ u(t)$ induces a zero-area cycle in shape space $ \mathcal{S} $ and does not produce any motion on the $z$ variable, see Fig.~\ref{fig:cont1}. 

\begin{figure}[htb!]
\begin{center}
\subfigure[]{
\includegraphics[width=6cm]{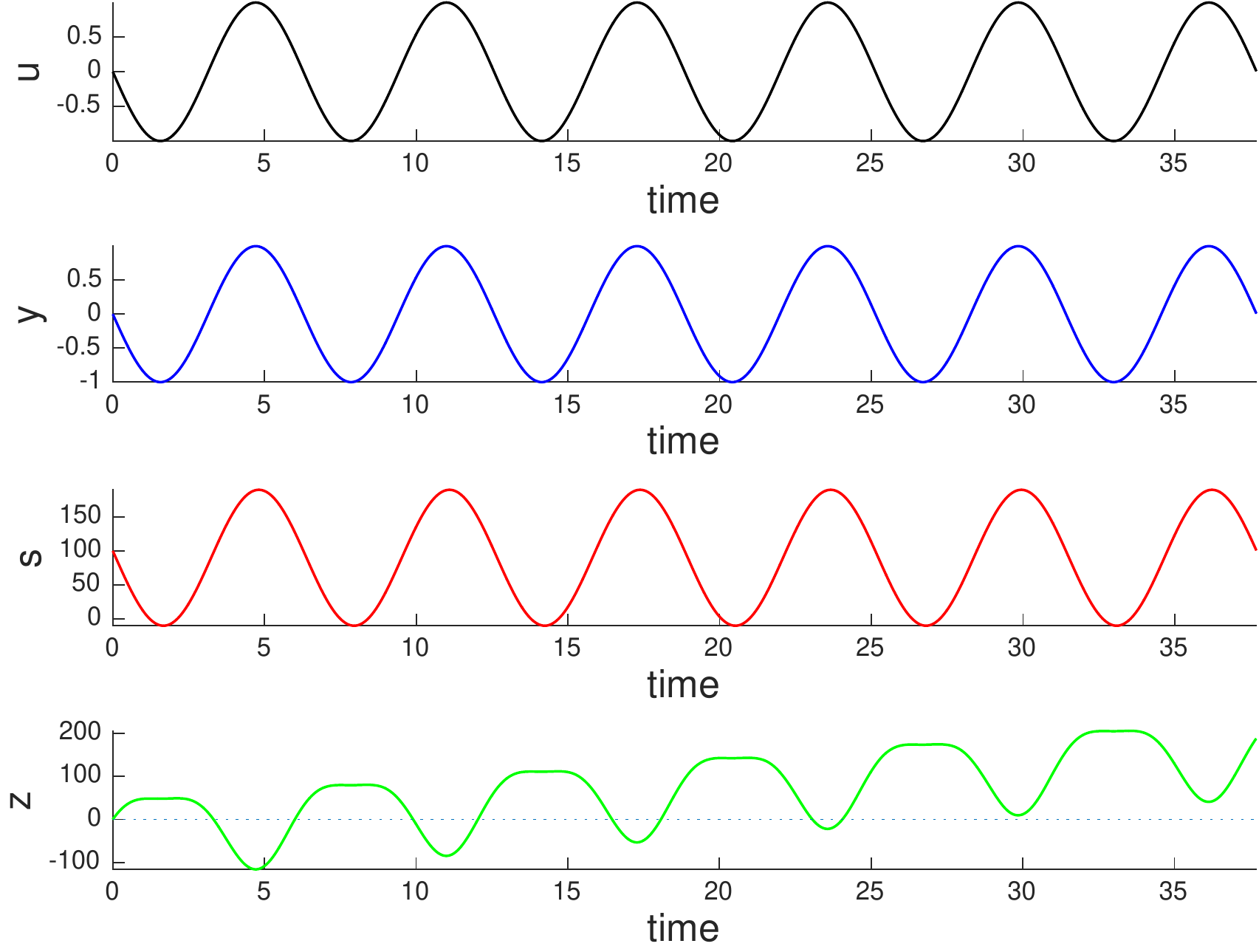}} $ \qquad $ 
\subfigure[]{
\includegraphics[width=6cm]{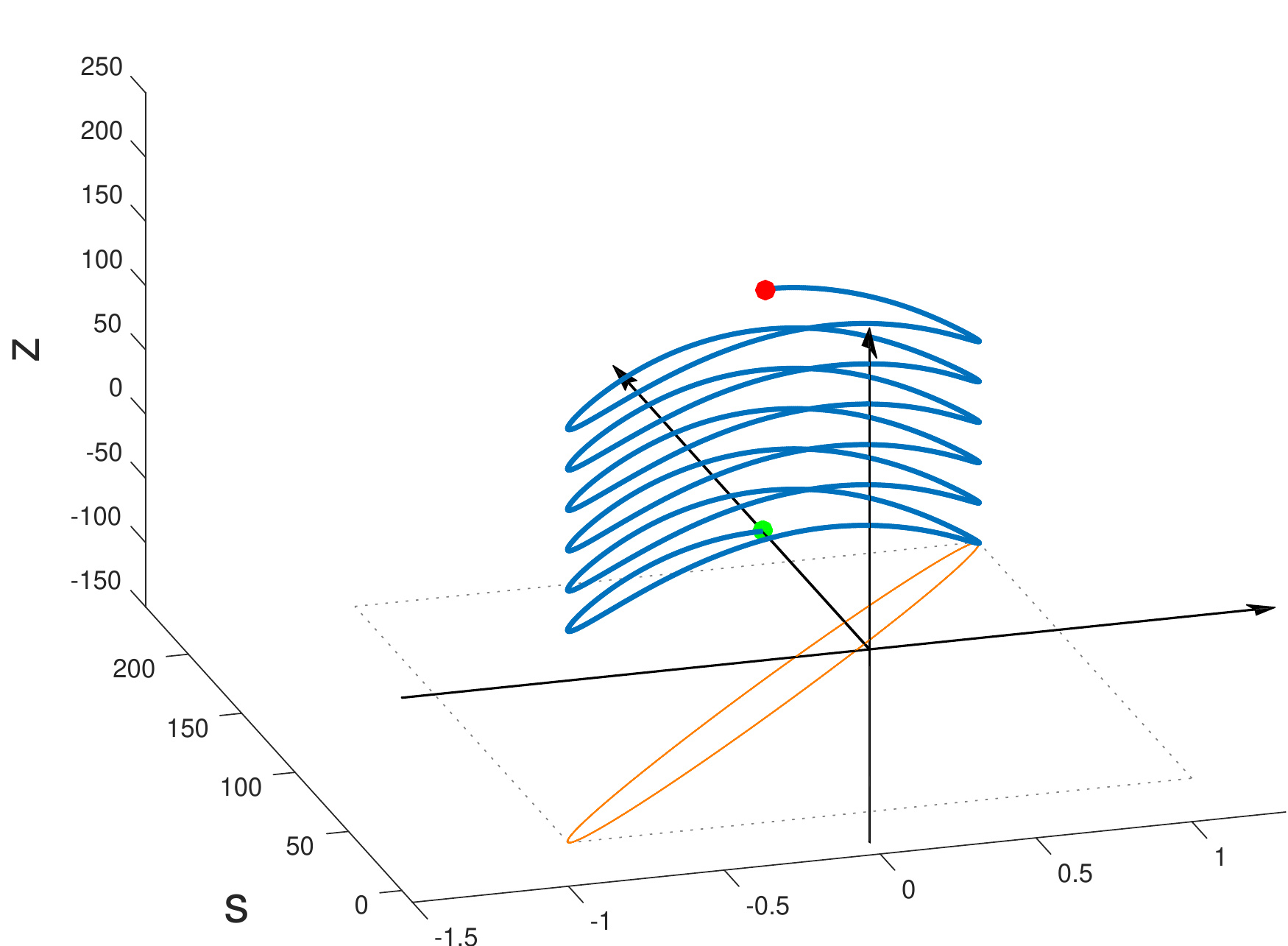}} 
\caption{Continuous-time stock trading model: adding a time delay in one of the ODEs (here that for the stock quote $ s$), is enough to induce a non-zero area in shape space and hence a geometric phase.}
\label{fig:cont2}
\end{center}
\end{figure}

In other words, the geometric phase described in the paper is a purely discrete-time phenomenon, with no continous-time counterpart. 
In order to obtain a similar effect in continuous-time, it is necessary to produce cyclic trajectories of non-zero area in the shape space $ (y, \, s )$.
This can be achieved for instance by introducing a delay in the dynamics of $ s$, representing a latency time in the response of the stock quote to buy/sell orders. 
Such delays are plausible if we focus on the very fast time scales preferred by high frequency traders.
If we replace \eqref{eq:continuous_model2} with 
\beq
\dot s(t) = r u(t - \tau) 
\label{eq:continuous_model2_delay}
\eeq
where $ \tau >0 $ is a time delay, then a periodic $ u$-trajectory induces a nonzero area in the plane $ \mathcal{S} $, and a net motion in the $ z$ variable is accumulated each time a cyclic trajectory is accomplished, see Fig.~\ref{fig:cont2}.

\begin{figure}[htb!]
\begin{center}
\subfigure[]{
\includegraphics[width=6cm]{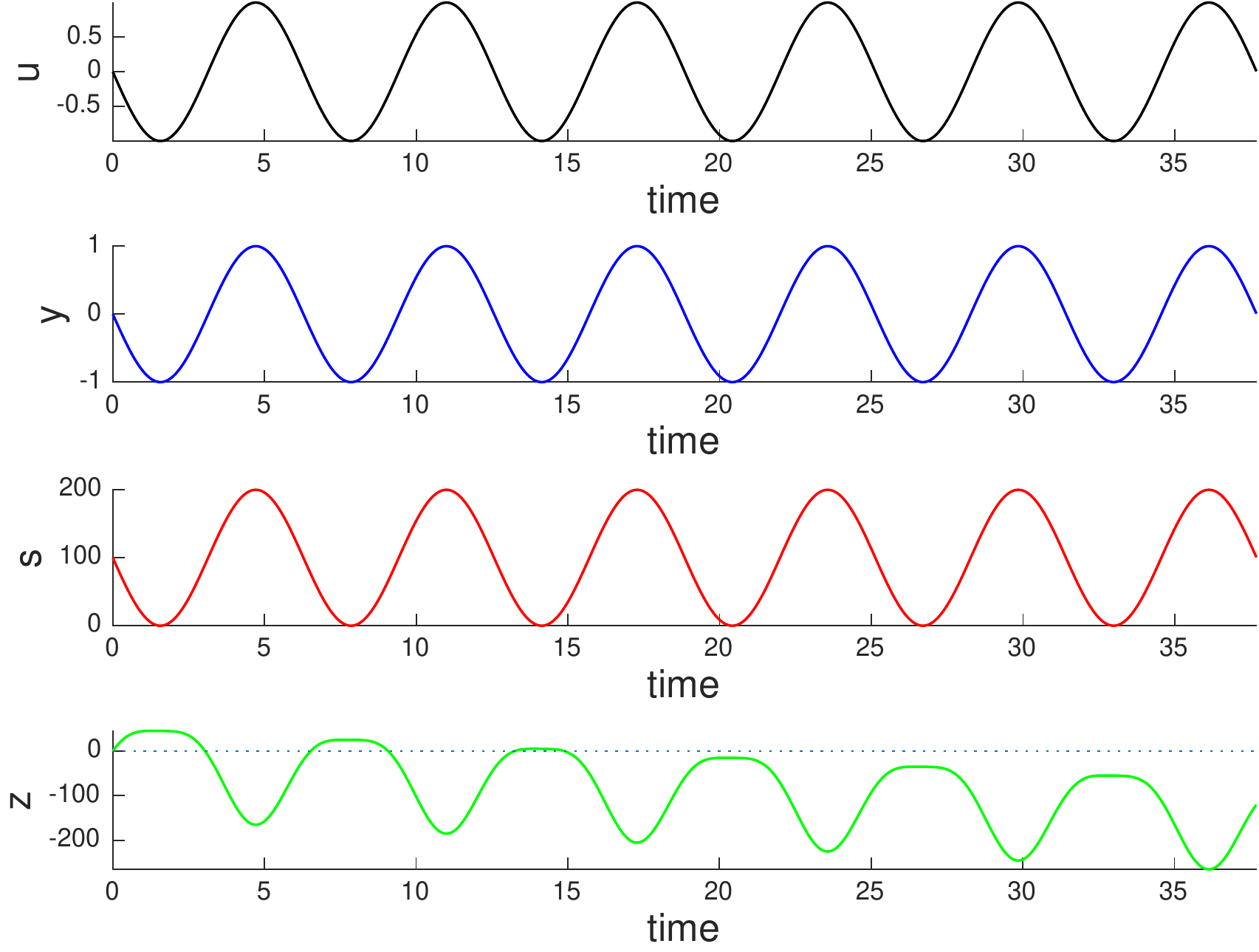}} $ \qquad $ 
\subfigure[]{
\includegraphics[width=6cm]{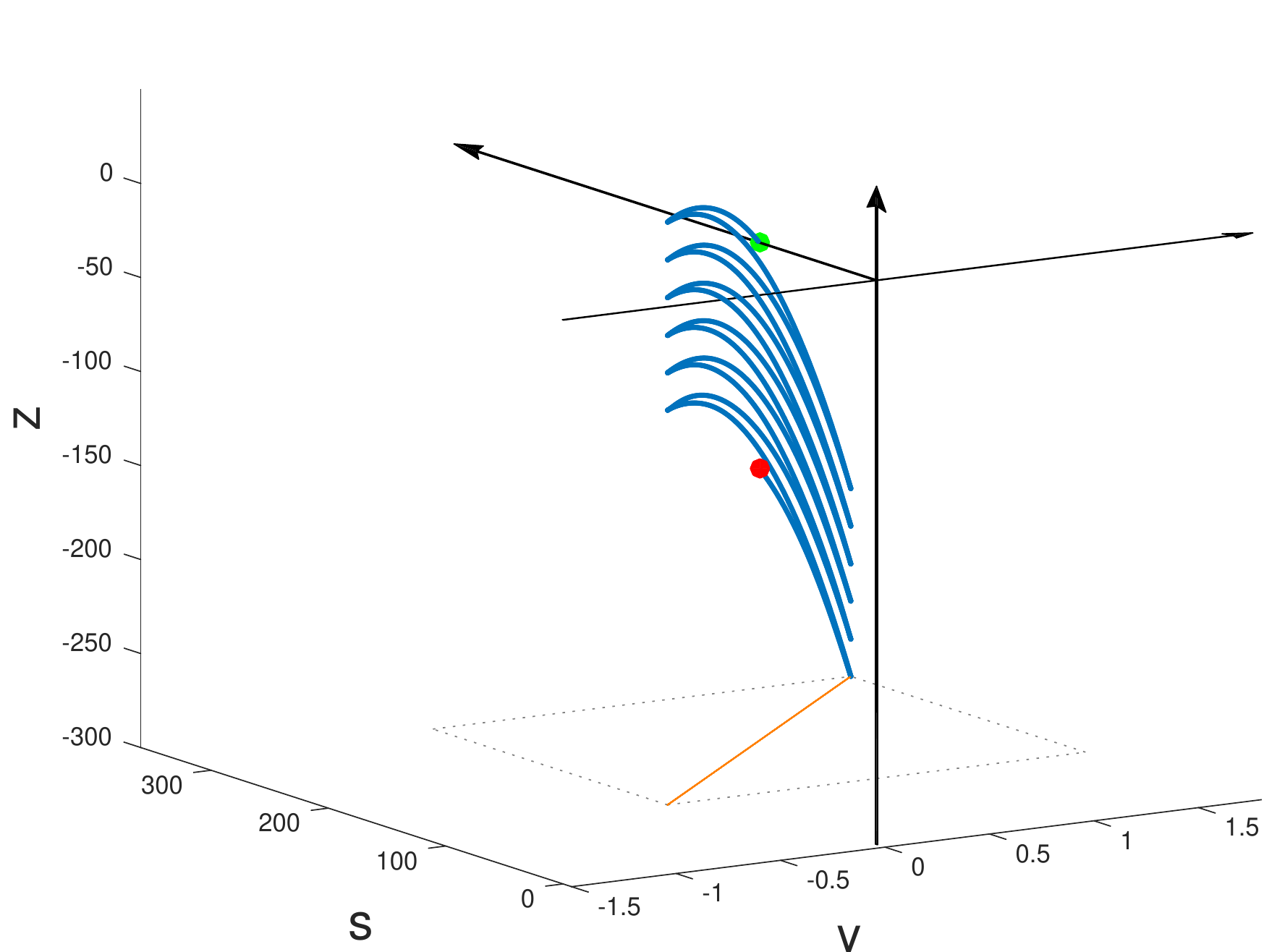}} 
\caption{Continuous-time stock trading model with quote spread and no time delay. Even though the cycle has zero area in shape space, a geometric phase is induced but it necessarily corresponds to a loss. The resulting ODE for the cash balance is discontinuous and uniqueness of the solution is not guaranteed.}
\label{fig:cont4}
\end{center}
\end{figure}

As in the discrete-time case, it is possible to include in the model a spread between the buy and sell prices. 
Considering a quote spread in the continuous-time model means replacing  \eqref{eq:continuous_model3} with 
\beq
\dot z (t) = \begin{cases} 
- a(t) u(t) & \text{if    } u(t) > 0 \\  
- b(t) u(t) & \text{if    } u(t) > 0 
\end{cases}
\label{eq:continuous_model3_spread}
\eeq
Since a spread reduces the profit margins, in the delay-free continuous-time model (i.e., eq.~\eqref{eq:continuous_model1}, \eqref{eq:continuous_model2} and \eqref{eq:continuous_model3_spread}), it is impossible to attain a positive cash flow out of cyclic operations only (unlike in the discrete-time case), see Fig.~\ref{fig:cont4}. 

This confirms the qualitatively different prediction given by discrete and continuous-time models. 
Notice in Fig.~\ref{fig:cont4} how a (negative) geometric phase is produced in spite of the zero-area of the cycle in shape space. This is however a pathological behavior, due to the discontinuity in the ODE \eqref{eq:continuous_model3_spread}, and comes at the price of loss of uniqueness of the solution of \eqref{eq:continuous_model3_spread} when $ u(t) $ crosses 0.

\begin{figure}[htb!]
\begin{center}
\subfigure[]{
\includegraphics[width=6cm]{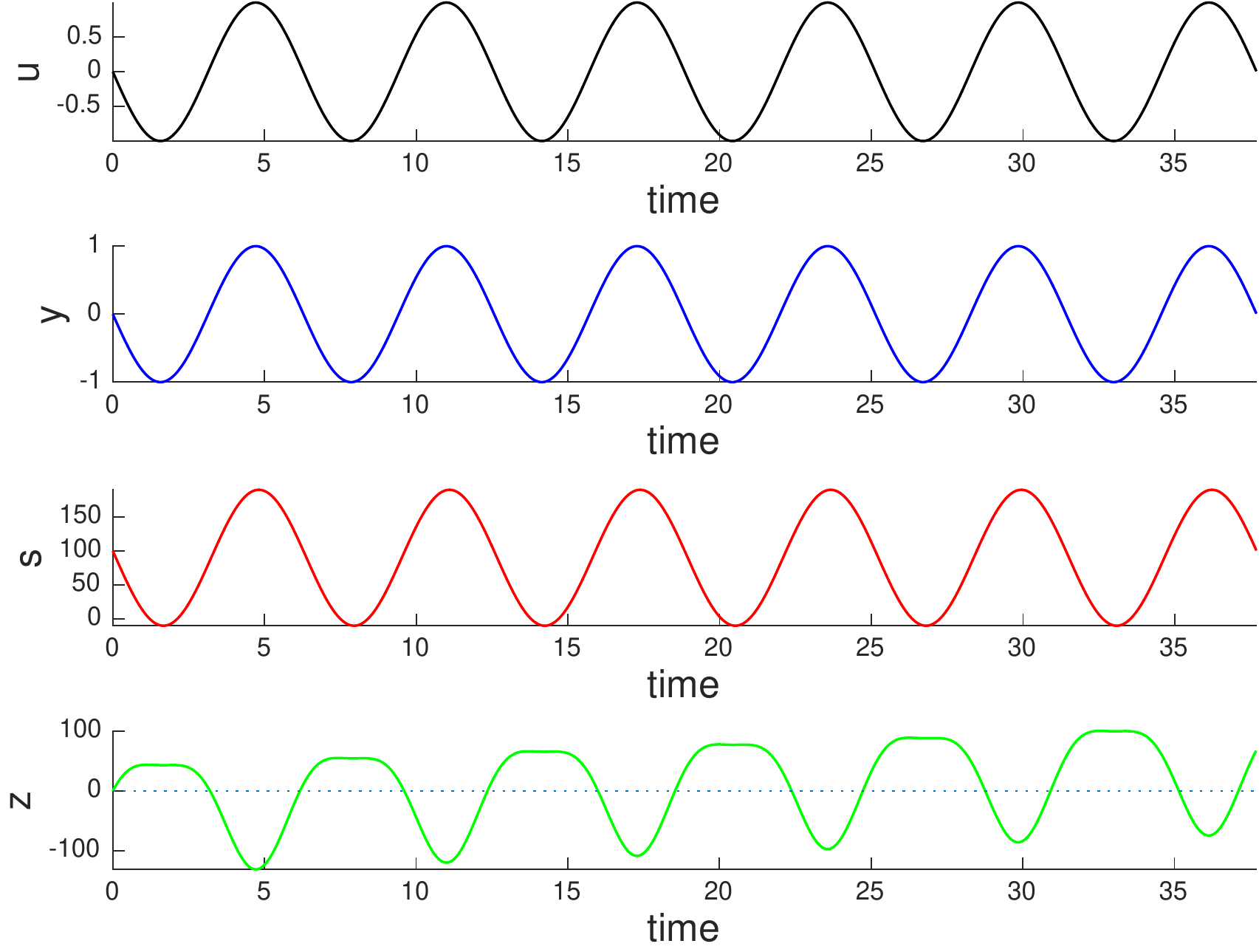}} $ \qquad $ 
\subfigure[]{
\includegraphics[width=6cm]{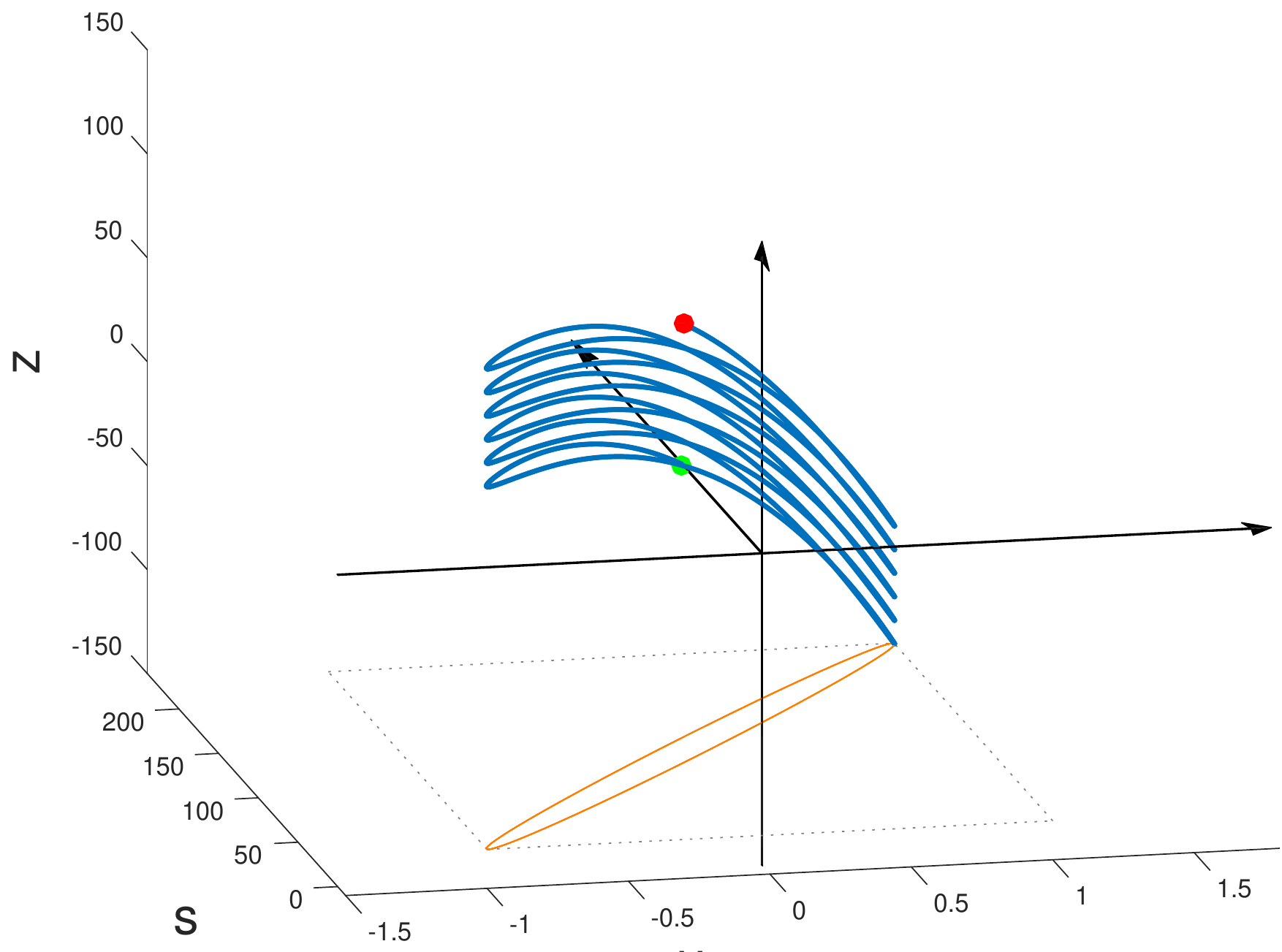}} 
\caption{Continuous-time stock trading model: when both a time delay and a quote spread are considered, the geometric phase can correspond to a profit, depending on the values of $ r$ and $ q$.}
\label{fig:cont3}
\end{center}
\end{figure}

When also a delay is added to the continuous-time model (i.e., eq.~\eqref{eq:continuous_model1}, \eqref{eq:continuous_model2_delay} and \eqref{eq:continuous_model3_spread} are considered), then a positive cash flow is again possible, depending on the numerical values of the parameters $ r$, $ \tau$, and $ q$, see Fig.~\ref{fig:cont3}.

%

\end{document}